\documentclass[twocolumn]{aastex63}
\usepackage[flushleft]{threeparttable}
\usepackage{graphicx}
\usepackage{amsmath}
\usepackage{natbib}

\newcommand{\GALEX}{\textit{GALEX}}

\graphicspath{{./}{figures/}}

\begin{document}

\title{The GALEX-PTF experiment: II. supernova progenitor radius and energetics via shock-cooling modeling}

\correspondingauthor{Noam Ganot}
\email{noam.ganot@weizmann.ac.il}

\author{Noam Ganot}
\affiliation{Department of Particle Physics and Astrophysics, Faculty of Physics, The Weizmann Institute of Science, Rehovot 76100, Israel}

\author[0000-0002-6786-8774]{Eran O. Ofek}
\affiliation{Department of Particle Physics and Astrophysics, Faculty of Physics, The Weizmann Institute of Science, Rehovot 76100, Israel}

\author[0000-0002-3653-5598]{Avishay Gal-Yam}
\affiliation{Department of Particle Physics and Astrophysics, Faculty of Physics, The Weizmann Institute of Science, Rehovot 76100, Israel}

\author[0000-0001-6753-1488]{Maayane T. Soumagnac}
\affiliation{Lawrence Berkeley National Laboratory, 1 Cyclotron Road, Berkeley, CA 94720, USA}
\affiliation{Department of Particle Physics and Astrophysics, Faculty of Physics, The Weizmann Institute of Science, Rehovot 76100, Israel}

\author{Jonathan Morag}
\affiliation{Department of Particle Physics and Astrophysics, Faculty of Physics, The Weizmann Institute of Science, Rehovot 76100, Israel}

\author[0000-0002-9038-5877]{Eli Waxman}
\affiliation{Department of Particle Physics and Astrophysics, Faculty of Physics, The Weizmann Institute of Science, Rehovot 76100, Israel}

\author[0000-0001-5390-8563]{Shrinivas R. Kulkarni}
\affil{Cahill Center for Astrophysics, California Institute of Technology, 1200 E. California Boulevard, Pasadena, CA 91125, USA}

\author[0000-0002-5619-4938]{Mansi M. Kasliwal}
\affil{Division of Physics, Mathematics, and Astronomy, California Institute of Technology, Pasadena, CA 91125, USA}

\author[0000-0002-0466-1119]{James Neill}
\affil{Division of Physics, Mathematics, and Astronomy, California Institute of Technology, Pasadena, CA 91125, USA}

\begin{abstract}
    The radius and surface composition of an exploding massive star, as well as the explosion energy per unit mass, can be measured using early ultraviolet (UV) observations of core-collapse supernovae (CC SNe). We present the results from a simultaneous \GALEX and Palomar Transient Factory (PTF) search for early UV emission from SNe. We analyze five CC SNe for which we obtained $NUV$ measurements before the first ground-based $R$-band detection. We introduce SOPRANOS, a new maximum likelihood fitting tool for models with variable temporal validity windows, and use it to fit the \citet{SapirWaxman2017} shock cooling model to the data. We report four Type II SNe with progenitor radii in the range of $R_*\approx600-1100R_\sun$ and a shock velocity parameter in the range of $v_{s*}\approx 2700-6000 \,\rm km\,s^{-1}$ ($E/M\approx2-8\times10^{50}\,\rm erg/M_\sun$) and one type IIb SN with $R_*\approx210R_\sun$ and  $v_{s*}\approx11000 \rm\, km\,s^{-1}$ ($E/M\approx1.8\times10^{51}\,\rm erg/M_\sun$). Our pilot GALEX/PTF project thus suggests that a dedicated, systematic SN survey in the $NUV$ band, such as the wide-field UV explorer \textit{ULTRASAT} mission, is a compelling method to study the properties of SN progenitors and SN energetics.
\end{abstract}

\section{Introduction}
A Core collapse Supernova (SN) explosion marks the end of life of a massive star. Although there is a wide range of evidence to support this, the details of the final stages of the evolution of such massive stars \citep[see, e.g.][and references within]{Langer2012} and the exact association between SN type and progenitor type are not firmly established \citep[see reviews by][]{Filippenko1997,SmarttEtAl2009,Smartt2015,GalYam2017}.

Most previous, existing and planned SN surveys such as the Palomar Transient Factory \citep[PTF;][]{LawEtAl2009,RauEtAl2009},  the Large Synoptic Survey Telescope \citep[LSST;][]{IvezicEtAl2019}, the All Sky Automated Survey for Supernovae \citep[ASAS SN;][]{ShappeeEtAl2014}, the Asteroid Terrestrial-impact Last Alert System \citep[ATLAS;][]{TonryEtAl2018}, Large Array Survey Telescope \citep[LAST;][]{OfekBenAmi2020} and the Zwicky Transient Facility \citep[ZTF;][]{Bellem2014} are limited to the visible-NIR wavelengths. Only a few SNe were detected by space telescopes in shorter wavelengths, often by coincidence \citep[e.g.][]{ArnettEtAl1989,SchmidtEtAl1993,CampEtAl2013,GezariEtAl2008,GezariEtAl2010,SchawinskiEtAl2008,SoderbergEtAl2008,OfekEtAl2010,ArcaviEtAl2011,GalYamEtAl2011,CaoEtAl2013,SoumagnacEtAl2020}. Theoretical models predict that the earliest emission from CC SNe is a burst of radiation occurring when the explosion shock breaks out of the stellar surface (the shock breakout flare). This flare is expected to have temperatures in the range of few to tens of eV, and its duration would be $R_*/c$, where $R_*$ is the stellar radius, i.e., a timescale of about 1 hour \citep{GrassbergEtAl1971, Colgate1974, Falk1978, KleinChevalier1978, EnsmanBurrows1992, MatznerMcKee1999, NakarSari2010, SapirEtAl2011, KatzEtAl2012,SapirEtAl2013}. At these temperatures, the UV emission is expected to be stronger than the one in the visible band. Following the breakout flare the expanding stellar envelope enters a shock cooling phase where it emits a fraction of the leftover explosion energy. This phase is better understood theoretically \citep[e.g.,][]{GrassbergEtAl1971,Chevalier1976,Chevalier1992,ChevalierFransson2008,NakarSari2010,RabinakWaxman2011,SapirWaxman2017}. According to those models it takes a few days for the photospheric temperature to cool to temperatures below 1eV and for the peak flux to move into the visible band. Fortunately, there were a few SNe which had some early near UV ($NUV$) measurements (by \GALEX, \citealt{GezariEtAl2008,SchawinskiEtAl2008}; and by Swift, \citealt{SoderbergEtAl2008}) which allowed to test these models at temperature $\gtrsim\rm 1eV$. \citet{RabinakWaxman2011} have demonstrated that early shock cooling phase data encodes information about the progenitor radius and surface composition, the SN explosion energy per unit mass and the line of sight extinction.

In a recent work, \cite{Goldberg2020} point out that models that follow the evolution of SNe II-P during the plateau phase suffer from an inherent degeneracy in determining the properties of the progenitors. This emphasizes the importance of early data and models which are valid in those time to be able to determine the SN progenitor properties. In this paper we present the results of the first systematic space-borne $NUV$ survey for SN and the fitting of the model of \citet{SapirWaxman2017} to the observations. A more comprehensive and sensitive survey will be conducted with the launch of the \textit{ULTRASAT} mission \citep{SagivEtAl2014} expected by 2024.

We present our observations on \S \ref{s:Observations}, describe the \citet{SapirWaxman2017} model and our \verb|SOPRANOS| fitting formalism in \S \ref{s:Analysis}, report our fit results in \S \ref{s:Results}, discuss them in \S \ref{s:Discussion} and conclude in \S \ref{S:Conculusion}.

\section{Observations}
\label{s:Observations}

\begin{table*}[t]
\caption{The core-collapse SNe detected by PTF during the GALEX/PTF experiment}
\centering
\begin{threeparttable}
\begin{tabular}{l l l l l l l l}
\hline
\hline
PTF name & J2000 RA & J2000 Dec & Redshift & Type & PTF discovery & $E_{B-V}$\tnote{a}  & NUV background \\
         &          &           &          &      & date          & [mag]       & [$\rm erg\,s^{-1}cm^{-2}\text{\AA}^{-1}$]\\
\hline
PTF12ffs & 14:42:07.33 & +09:20:29.8 & 0.0511 & SN II\tnote{b}  & June 10, 2012 & 0.025 & $2.24\times10^{-16}$\\
PTF12fhz & 15:18:20.09 & +10:56:42.7 & 0.0987 & SN IIb          & June 12, 2012 & 0.038 & $6.09\times10^{-18}$\\
PTF12fkp & 14:46:54.81 & +10:31:26.4 & 0.12   & SN II-L         & June 14, 2012 & 0.026 & $1.45\times10^{-17}$\\
PTF12ftc & 15:05:01.88 & +20:05:54.6 & 0.0732 & SN II-P         & June 19, 2012 & 0.036 & $4.24\times10^{-17}$\\
PTF12glz & 15:54:53.04 & +03:32:07.5 & 0.0799 & SN IIn          & July 7, 2012  & 0.132 & $4.84\times10^{-17}$\\
PTF12gnt & 17:27:47.30 & +26:51:22.1 & 0.029  & SN II-P         & July 9, 2012  & 0.047 & $1.20\times10^{-15}$\\
\hline
PTF12fes & 16:00:35.13 & +15:41:03.5 & 0.0359 & SN Ib           & June 9, 2012  & 0.038 & $4.24\times10^{-17}$\\
PTF12fip & 15:00:51.04 & +09:20:25.1 & 0.034  & SN II-P         & June 12, 2012 & 0.030 & $4.66\times10^{-16}$\\
PTF12frn & 16:22:00.16 & +32:09:38.9 & 0.136  & SN IIn          & June 18, 2012 & 0.021 & $5.08\times10^{-16}$\\
PTF12gcx & 15:44:17.32 & +09:57:43.1 & 0.045  & SN II\tnote{c}  & June 26, 2012 & 0.054 & $5.18\times10^{-16}$ \tnote{d}\\
\hline
\end{tabular}
\begin{tablenotes}
             \item[] For the four SNe below the line, no matching $NUV$ transient was identified in the \GALEX images.
            \item[a] The Galactic extinction according to \cite{SchlegelEtAl1998} maps.
            \item[b] A bright SN II with a light curve intermediate between SNe II-P and II-L
            \item[c] A bright SN II with a very long rise time, similar to SN 1998A \citep{PastorelloEtAl2005}, SN 2000cb \citep{KleiserEtAl2011} and SNe 2005ci and 2005dp \citep{ArcaviEtAl2012}. see also \citet{TaddiaEtAl2015}.
            \item[d] The PTF12gcx background was calculated using only \GALEX data measurements prior to the SN detection by PTF,   since the $NUV$ transient  could not be localized.
\end{tablenotes}
\end{threeparttable}
\label{tab:SNtable}
\end{table*}

Between 2012 May 24 and 2012 July 28 we monitored about 600\,deg$^{2}$ of high galactic latitude sky using the {\it GALEX} satellite \citep{MartinEtAl2005} and the Palomar Transient Factory \citep[PTF;][]{LawEtAl2009,RauEtAl2009}.
The observations were conducted with the \GALEX-$NUV$ filter and the PTF Mould $R$ band.
This \GALEX-PTF experiment is further described in \cite{GanotEtAl2016}.
During the experiment ten core collapse Supernovae were detected in the PTF \textit{R}-band observations. For six SNe out of the ten a matching transient was detected in the \GALEX $NUV$ observations. In this paper we analyze the light curves of the SNe detected in the experiment. The SNe detected are listed in Table \ref{tab:SNtable} as appeared in \citet{GanotEtAl2016}.

\subsection{\textit{GALEX} data}
\begin{table}[h]
\caption{PTF12ffs \GALEX $NUV$ data}
\centering
\begin{threeparttable}
\begin{tabular}{l l l l}
\hline
\hline
MJD	        & CPS\tnote{a}       & CPSERR\tnote{b} & $M_{AB}$\\
$[$days$]$  &           &        &       \\
\hline
56073.411	& 0.664	    & 0.148  & 20.52\\
56073.412	& 1.214	    & 0.166  & 19.87\\
56076.151	& 0.822	    & 0.154  & 20.29\\
56076.151	& 1.267	    & 0.180  & 19.82\\
56078.890	& 2.851	    & 0.282  & 18.94\\
... \\
\hline
\end{tabular}
\begin{tablenotes}
    \item The full \GALEX data for the events reported in the paper is available in the electronic version.
    \item[a] counts s$^{-1}$
    \item[b] counts s$^{-1}$ $1\sigma$ error
\end{tablenotes}
\end{threeparttable}
\label{tab:GALEXdata}
\end{table}
During our experiment, \GALEX \citep{MartinEtAl2005} was operating in scanning mode\footnote{\url{http://galex.stsci.edu/gr6/?page=scanmode}} and its NUV camera \citep{MorrisseyEtAl2007} observed strips in the sky in a drift-scan mode with an effective average exposure time of about  80s, reaching a $NUV$ 5$\sigma$ limiting magnitude of 20.6 mag AB. Each strip was visited once every three days. The \GALEX $NUV$ photometry was measured using custom aperture photometry routines \citep{Ofek2014} and is described in details in \citet{GanotEtAl2016}. Some of the SNe appear in more than one strip sub-scan leading to two adjacent visits within the \GALEX three day cadence. Some of the scans were lost due to failed downlink and image corruption leading to missing points on the light curves. The \GALEX photometry observations are listed in Table \ref{tab:GALEXdata}, and plotted together with a shock cooling fit of each SN (see \S\ref{s:Results}; Figs. \ref{fig:12ffs_LC}, \ref{fig:12gnt_LC}, \ref{fig:12fkp_LC}, \ref{fig:12ftc_LC}, \ref{fig:12fhz_LC}).

\subsection{PTF data}
\begin{table}[h]
\caption{PTF12ffs PTF $R$ data}
\centering
\begin{threeparttable}
\begin{tabular}{l l l l l}
\hline
\hline
MJD         & counts    & dcounts &zero   & $M_{AB}$\\
$[$days$]$  &           &         &point  &         \\
\hline
56077.192   &     -4.7  &   85.3 & 27.000 &(21)\tnote{a}\\
56077.227   &   -117.9  &   75.0 & 27.000 &(21.93)\tnote{a}\\
56077.257   &     35.8  &   74.4 & 27.000 &23.12\\
56081.188   &   1126.8  &  198.5 & 27.000 &19.37\\
56081.220   &   1262.9  &  216.6 & 27.000 &19.25\\
... \\
\hline
\end{tabular}
\begin{tablenotes}
    \item The full PTF data for the events reported in the paper is available in the electronic version.
    \item [a] Negative count values cannot be converted to magnitudes. Instead of magnitudes we report  $3\sigma$ limit magnitudes for these measurements.
\end{tablenotes}
\end{threeparttable}
\label{tab:PTFdata}
\end{table}

The Palomar Transient Factory (PTF), using the 48 inch Oschin Schmidt Telescope at Palomar Observatory equipped with a CCD mosaic, with a field of view of $7.26\,\rm deg^2$, was scanning a similar sky patch as \GALEX in its Mould $R$-band, reaching an $R$-band limiting magnitude of about 21 mag AB with a cadence of two days, weather permitting. During the experiment PTF discovered ten core collapse SNe, listed in Table \ref{tab:SNtable}.
The photometry was extracted using a point-spread function (PSF) fitting routine \citep{SullivanEtAl2006,FirthEtAl2015} applied after image subtraction that was done using the \citet{AlardLupton1998} algorithm. The PTF data reduction is described in \citet{LaherEtAl2014}, while the photometric calibration and magnitude system are described by \citet{OfekEtAl2012a}. The PTF photometry observations are listed in Table \ref{tab:PTFdata}, and plotted with the SN individual fits  (see \S\ref{s:Results}; Figs. \ref{fig:12ffs_LC}, \ref{fig:12gnt_LC}, \ref{fig:12fkp_LC}, \ref{fig:12ftc_LC} and \ref{fig:12fhz_LC}).

\subsection{Spectra}
\begin{table}[t]
\caption{Spectroscopy measurement for GALEX/PTF SNe}
\centering
\begin{threeparttable}
\begin{tabular*}{0.9\linewidth}{l l l l l l}
\hline
\hline
SN name & Telescope & Instrument & Date \\
\hline
PTF12ffs & Keck I & LRIS  & 2012 Jun 16 \\
PTF12ffs & P200 & DBSP &  2012 Jul 21 \\
\hline
PTF12fhz & Keck II & DEIMOS & 2012 Jul 16 \\
PTF12fhz & Keck I & LRIS & 2012 Jul 18 \\
PTF12fhz & Keck I & LRIS & 2012 Jul 19 \\
PTF12fhz & Keck I & LRIS & 2012 Sep 19 \\
PTF12fhz & Keck I & LRIS & 2013 Feb 09 \\
\hline
PTF12fkp & Keck I & LRIS & 2012 Jul 15 \\
PTF12fkp & Keck I & LRIS & 2012 Jul 15 \\
PTF12fkp & Keck I & LRIS & 2012 Aug 19 \\
\hline
PTF12ftc & Keck I & LRIS & 2012 Jul 15 \\
PTF12ftc & Keck I & LRIS & 2012 Jul 15 \\
PTF12ftc & P200 & DBSP &  2012 Jul 27 \\
\hline
PTF12glz & P200 & DBSP & 2012 Jul 15\\
PTF12glz & P200 & DBSP & 2012 Jul 26\\
PTF12glz & Keck I & LRIS & 2013 Feb 09\\
PTF12glz & Keck I & LRIS & 2013 May 09\\
\hline
PTF12gnt & APO & DIS & 2012 Jul 16 \\
PTF12gnt & Keck I & LRIS & 2012 Jul 16 \\
PTF12gnt & P200 & DBSP & 2012 Jul 21 \\
PTF12gnt & P200 & DBSP & 2012 Jul 26 \\
PTF12gnt & Keck I & LRIS & 2012 Aug 18 \\
\hline
\end{tabular*}
\begin{tablenotes}
    \item Keck I LRIS - Keck I 10 m telescope Low Resolution Imaging Spectrometer 
    \item Keck II DEIMOS - Keck II 10 m telescope DEep Imaging Multi-Object Spectrograph 
    \item APO DIS - Apache Point Observatory 3.5 m telescope Dual Imaging Spectrograph
    \item P200 DBSP - Palomar 200-inch Hale telescope Double Beam Spectrograph
\end{tablenotes}
\end{threeparttable}
\label{tab:Spectable}
\end{table}

\begin{figure}
\includegraphics[width=\linewidth]{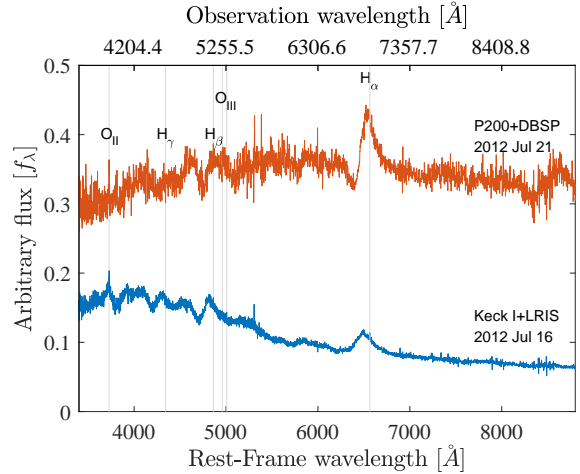}
\caption{Spectra of PTF12ffs obtained from various observatories (see Table \ref{tab:Spectable}). We used $H_\alpha$, $H_\beta$, $H_\gamma$, $O_{\rm II}$ and $O_{\rm III}$ common galaxy lines (plotted) to measure a red shift of 0.0511.}
\label{fig:12ffs_spec}
\end{figure}

\begin{figure}
\includegraphics[width=\linewidth]{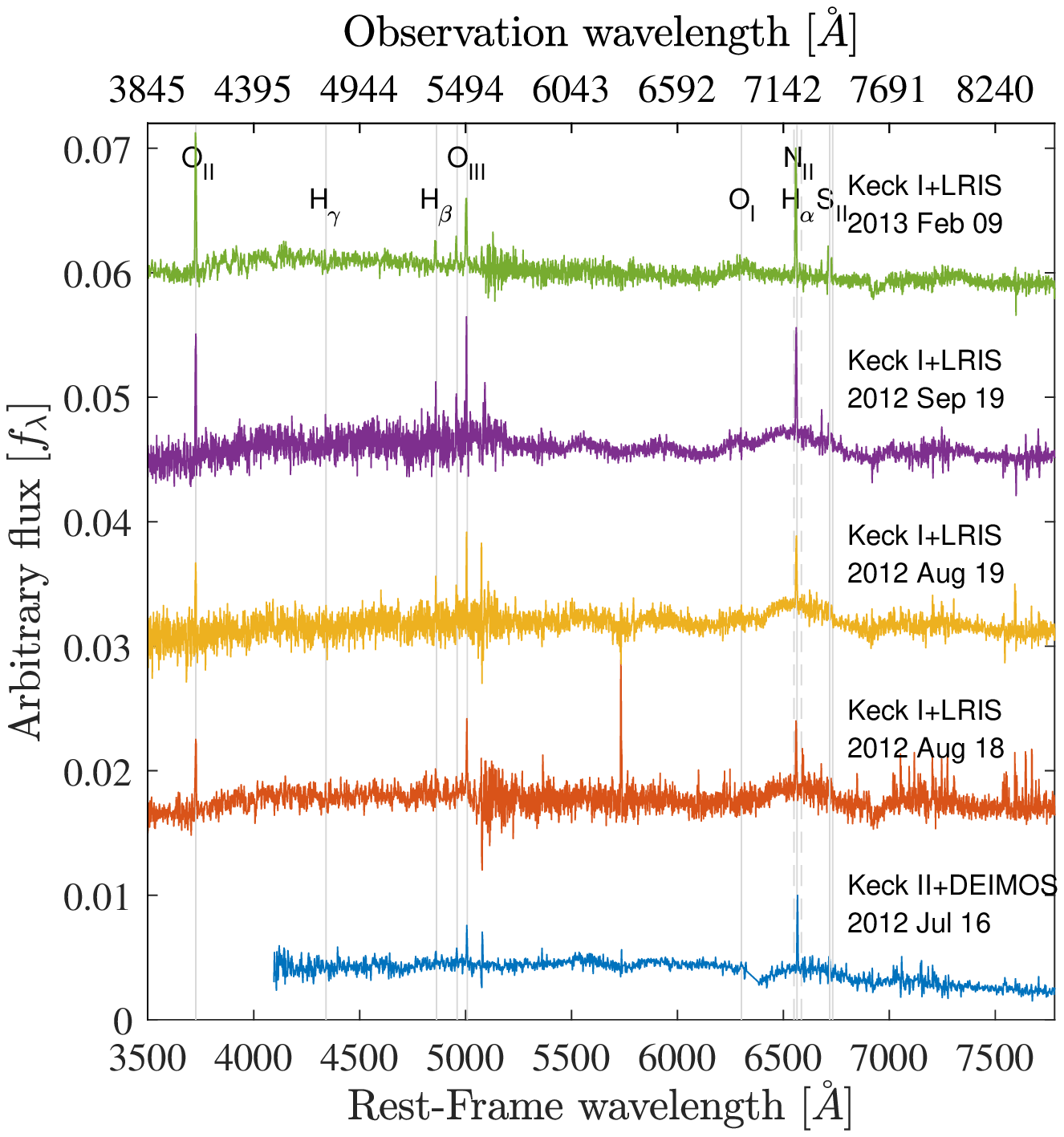}
\caption{Spectra of PTF12fhz (see Table \ref{tab:Spectable}). We used $H_\alpha, H_\beta, H_\gamma, O_{\rm I}, O_{\rm II}, O_{\rm III}, N_{\rm II}$ and $S_{\rm II}$ common galaxy lines (plotted) to measure a red shift of 0.0987.}
\label{fig:12fhz_spec}
\end{figure}

\begin{figure}
\includegraphics[width=\linewidth]{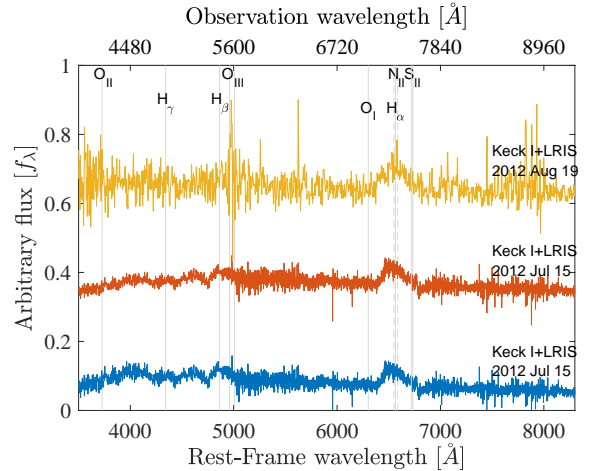}
\caption{Spectra of PTF12fkp (see Table \ref{tab:Spectable}). We used $H_\alpha, H_\beta, H_\gamma, O_{\rm I}, O_{\rm II}, O_{\rm III}, N_{\rm II}$ and $S_{\rm II}$ common galaxy lines (plotted) to measure a red shift of 0.12. The Keck I LRIS spectrum from Aug 19$^{th}$ was binned to reduce noise.}
\label{fig:12fkp_spec}
\end{figure}

\begin{figure}
\includegraphics[width=\linewidth]{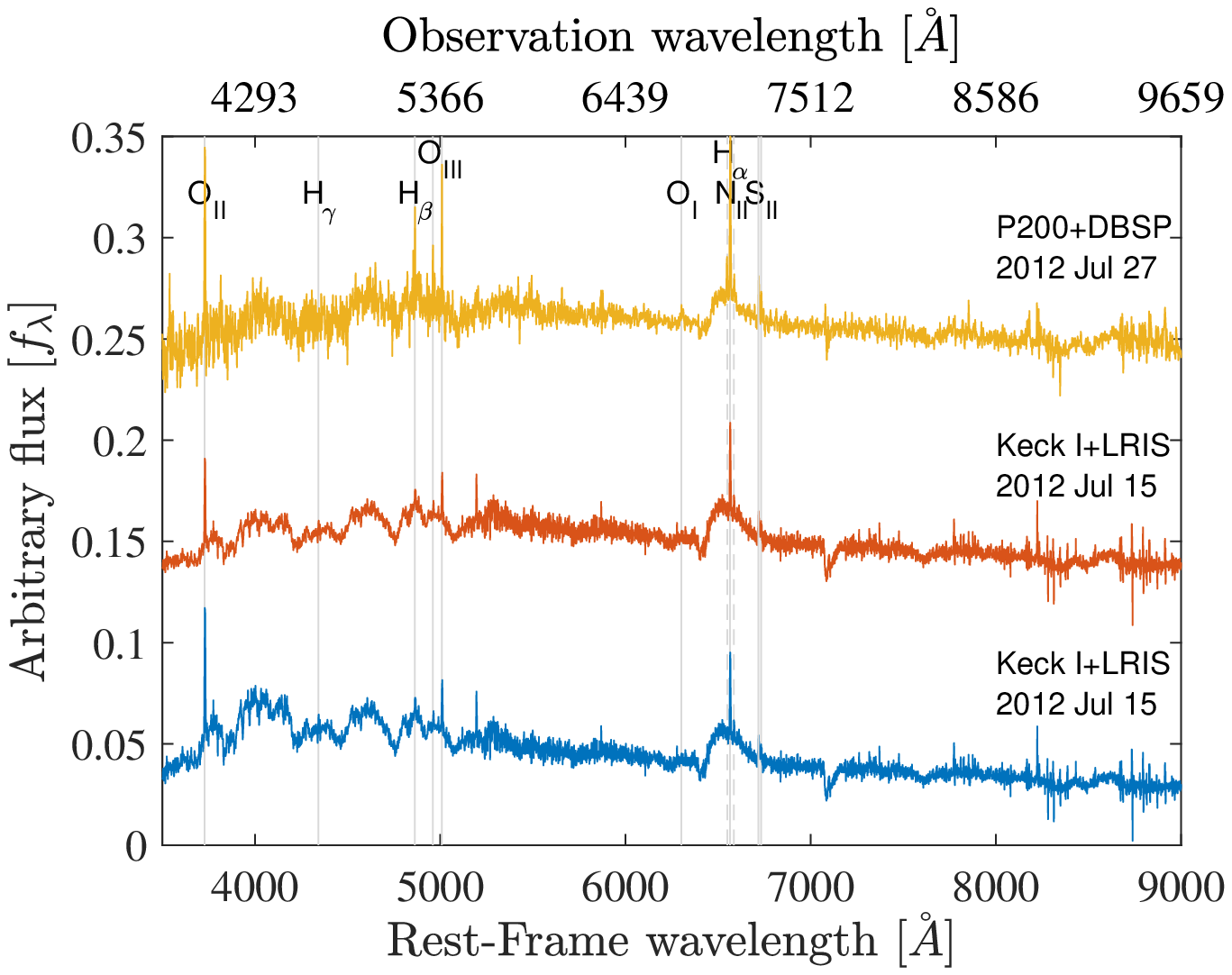}
\caption{Spectra of PTF12ftc (see Table \ref{tab:Spectable}). We used $H_\alpha, H_\beta, H_\gamma, O_{\rm I}, O_{\rm II}, O_{\rm III}, N_{\rm II}$ and $S_{\rm II}$ common galaxy lines (plotted) to measure a red shift of 0.0732.}
\label{fig:12ftc_spec}
\end{figure}

\begin{figure}
\includegraphics[width=\linewidth]{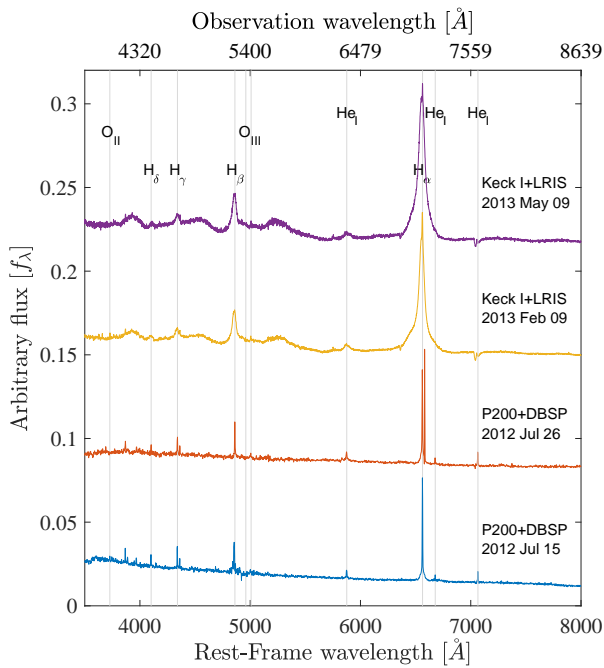}
\caption{Spectra of PTF12glz (see Table \ref{tab:Spectable}). We used $H_\alpha, H_\beta, H_\gamma, H_\delta, He_{\rm I}, O_{\rm II}$, and $O_{\rm III}$ lines (plotted) to measure a red shift of 0.0799.}
\label{fig:12glz_spec}
\end{figure}

\begin{figure}
\includegraphics[width=\linewidth]{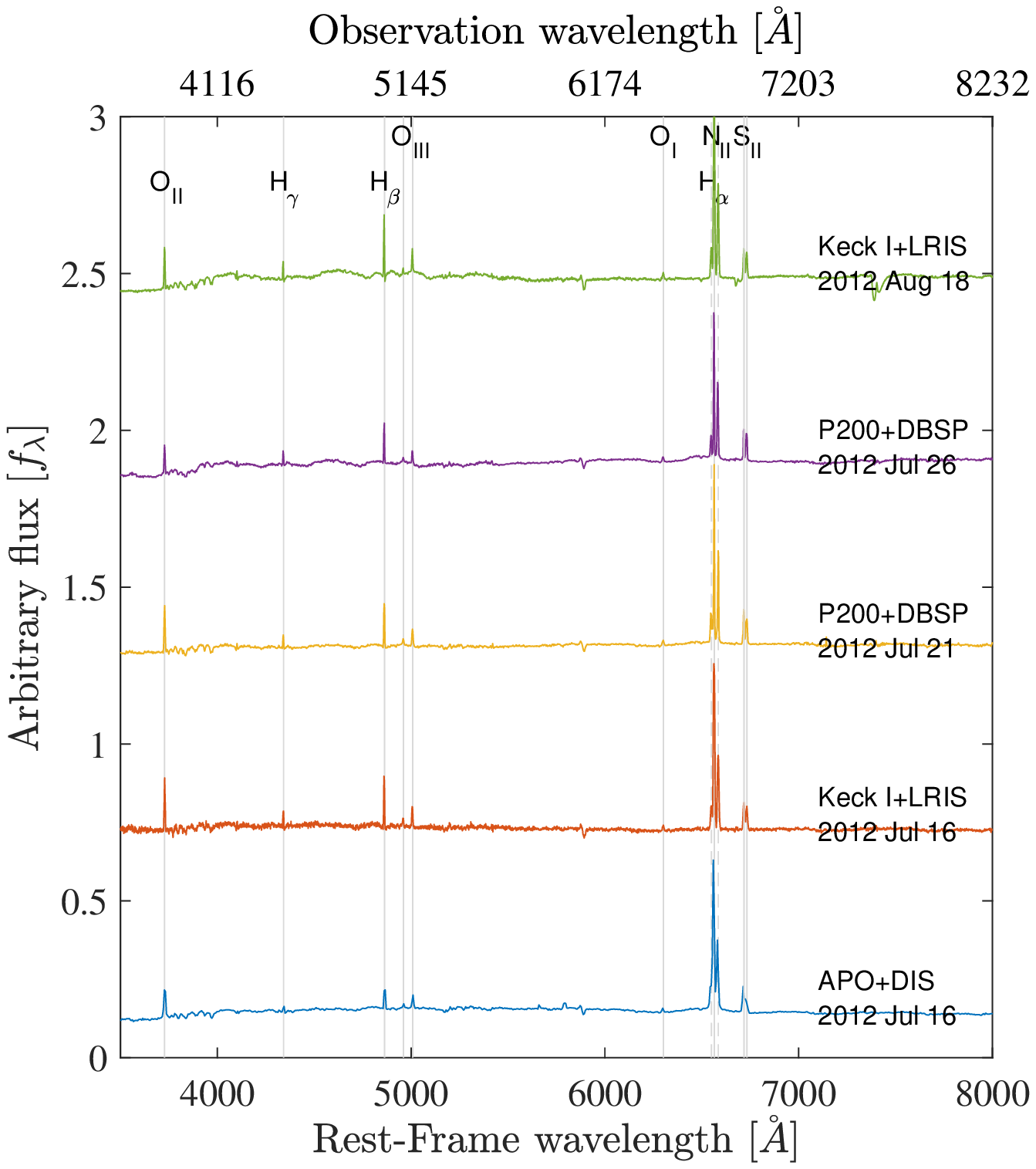}
\caption{Spectra of PTF12gnt (see Table \ref{tab:Spectable}). We used $H_\alpha, H_\beta, H_\gamma, O_{\rm I}, O_{\rm II}, O_{\rm III}, N_{\rm II}$ and $S_{\rm II}$ common galaxy lines (plotted) to measure a red shift of 0.029.}
\label{fig:12gnt_spec}
\end{figure}

Following the PTF transient detection several spectra were taken using the Keck I 10 m telescope Low Resolution Imaging Spectrometer \citep[LRIS;][]{OkeEtAl1995}, the Keck II 10 m telescope DEep Imaging Multi-Object Spectrograph \citep[DEIMOS;][]{FaberEtAl2003}, the Apache Point Observatory 3.5 m telescope Dual Imaging Spectrograph\footnote{\url{http://www.apo.nmsu.edu/arc35m/Instruments/DIS/}}, and DBSP, the Palomar 200-inch Hale telescope Double Beam Spectrograph\footnote{\url{http://www.astro.caltech.edu/palomar/observer/200inchResources/dbspoverview.html}}. The spectra are listed in Table \ref{tab:Spectable} and shown in Figures \ref{fig:12ffs_spec}, \ref{fig:12fhz_spec} ,\ref{fig:12fkp_spec}, \ref{fig:12ftc_spec}, \ref{fig:12glz_spec} and \ref{fig:12gnt_spec}. All observations are made public via WiseREP\footnote{\url{https://wiserep.weizmann.ac.il/}} \citep{YaronGalyam2012}. All spectra include strong absorption lines indicating that the photosphere was already recombined at the epochs the spectra were taken.

\section{Analysis}\label{s:Analysis}
We fit the observed $NUV$ and \textit{R}-band light curves with shock cooling models described in \citet[hereafter RW11]{RabinakWaxman2011} with the modifications of \citet[hereafter SW17]{SapirWaxman2017} and Morag et al. (2020, in prep., hereafter MSW20). All these models describe the shock cooling phase of a SN and do not treat the interaction between the SN radiation and a circumstellar material (CSM). Therefore they can be applied only for SNe with no evidence for CSM interaction in their spectra.
The free parameters we fit are the progenitor radius ($R_*$), shock velocity parameter ($v_{s*}$), ejecta mass ($M_{\rm ej}$), progenitor envelope structure parameter ($f_\rho$), reference time ($t_{\rm ref}$) and the extinction ($E_{\rm B-V}$). The shock velocity parameter $v_{s*}$ is defined by the Gandel'Man-Frank-Kamenetskii-Sakurai self-similar solutions \citep{GandelManEtAl1956,Sakurai1960} for shockwave propagation for a thin shell near the edge of the progenitor, $\delta \equiv (R_* - r)/R_*\ll 1$:
\begin{equation}
    v_{\rm sh}(\delta)=v_{s*}\delta^{-\beta n},
\end{equation}
where $r$ is the shell initial radius, $n = 3/2 (3)$ is the polytropic index for convective (radiative) envelopes \citep{MatznerMcKee1999}, $v_{\rm sh}(\delta)$ is the shock velocity of the shell $\delta$ and $\beta = 0.191 (0.186)$. $f_\rho$ is defined by the progenitor density profile near its surface,
\begin{equation}
    \rho_0 = f_\rho\overline{\rho}_0\delta^n,
\end{equation}
where $\overline{\rho}_0=3M_{\rm ej}/(4\pi R_*^3)$. The reference time ($t_{\rm ref}$) is the model zero time (see discussion in \S\ref{S:model} below).
A problem with the RW11 and SW16 models is that the time range during which they are valid depends on the physical parameters we are attempting to fit. This means that not all the observations can be used, and that the specific observations that can be used (and hence the number of degrees of freedom) depend on the model parameters. Furthermore, this also means that there are regions in the phase space which we cannot constrain. This problem is partially alleviated with the introduction of the MSW20 model extension, which is valid from the earliest time relevant to our observations, but still affects the model validity for the late data points (See for example the model validity range  in Figure \ref{fig:12gnt_LC}; each model is plotted only within its validity period).
We note that this is an important limitation of the models and this fact is sometime ignored in the literature and may lead to biased results \citep[see the discussion in][]{RubinEtAl2017}.

Comparison between RW11 and other shock-cooling models, for example, those of \citet{NakarSari2010} was reported by \citet{SapirWaxman2017}. In particular, RW11 find that the temperature $T\sim t^{-0.45}$ while  \citet{NakarSari2010} consistently find $T\sim t^{-0.4} - t^{-0.6}$ during the relevant phases. Since the temporal dependency of both models is similar, the low cadence of our data does not allow us to distinguish between the two. Here, we use the model of SW17, an extension of RW11 which allows us to use late data points extending beyond the original RW11 validity range (see below), as well as the extension of MSW20, which allows us to utilize the light curve early data points, before RW11 and SW17 become valid.

\subsection{Model Description and Limitations}\label{S:model}
RW11 derive the radius and temperature of the photosphere, in a thin layer near the edge of the progenitor $\delta=1-r/R_*\ll1$, where $r$ is the pre-explosion radius of the shell. Here, we use their main results but we limit the discussion to red supergiants with $n=3/2$. The photosphere temperature and bolometric luminosity are given by (only $n=3/2$ values are given, see RW11 and SW17 for $n=3$ values)
\begin{equation}\label{eq:RW11}
\begin{split}
T_{\rm ph,RW} & = 1.61\left(\frac{v_{s*,8.5}^2t_d^2}{f_\rho M_0 \kappa_{0.34}}\right)^{\epsilon_1}\frac{R_{13}^{1/4}}{\kappa_{0.34}^{1/4}}t_d^{-1/2}\quad\rm eV, \\
L_{\rm RW} & = 2.0\times10^{42}\left(\frac{v_{s*,8.5}t_d^2}{f_\rho M_0 \kappa_{0.34}}\right)^{-\epsilon_2}\frac{v_{s*,8.5}^2R_{13}}{\kappa_{0.34}}\quad\rm erg\,s^{-1},
\end{split}
\end{equation}
where $R_*=10^{13}R_{13}\,\rm cm$, $M_{\rm ej}=10^0M_0M_\sun$, $v_{s*}=10^{8.5}v_{s*,8.5}\,\rm cm\,s^{-1}$, $\kappa=0.34\kappa_{0.34}\,\rm cm^2g^{-1}$, $t_d$ is the time elapsed since $t_{\rm ref}$ in days and the power-laws are $\epsilon_1$=0.027 and $\epsilon_2$=0.086. We have used a constant opacity of $\kappa_{0.34}=1$ for all the calculations in this paper.

Some papers confuse the model parameter $t_{\rm ref}$, with the SN explosion time $t_{\rm exp}$ or with the shock break out time $t_{\rm BO}$. In general, $t_{\rm ref}\neq t_{\rm BO}\neq t_{\rm exp}$, since RW11 is focused on late times and therefore ignores the shells initial radius, and their acceleration to their final velocity
\begin{equation}\label{eq:r=vt}
r(\delta_m,t)=v_f(\delta_m)t,
\end{equation}
where $\delta_m$ is a notation for the layer for which a fraction $\delta_m$ of the progenitor mass lies ahead of. These approximations are justified when dealing with the shock cooling phase of the explosion, after a significant expansion of the ejecta. However, if we try to put some physical meaning to $t_{\rm ref}$, the time $t_{d}=0$, Eq. \ref{eq:r=vt} will give us $r(\delta_m,0)=0$ which is clearly wrong. Since each layer starts at a finite radius $r_0(\delta_m)$, the time $t_{\rm ref}$ (i.e., $t_{d}=0$), is always earlier than $t_{\rm BO}$, the time the shock breaks out from the progenitor surface. We can estimate the time difference between $t_{\rm BO}$ and $t_{\rm ref}$ by calculating the time it takes a layer to artificially expand from 0 to $r_0$ at a velocity $v_f$, the final velocity of the shell after its acceleration phase is completed. This time is estimated to be of order of 1 day, for large stars $R_*\approx1000R_{\sun}$. Therefore we can conclude that $t_{\rm ref}$ is about 1 day before the shock breakout.

Both the RW11 and SW17 models become valid only when the ejecta shells have expanded significantly, reaching their terminal velocity, and when the photosphere penetrates deeper than the layer at which the initial breakout took place. These two conditions are met at
\begin{equation} \label{eq:t_min}
t>t_{\rm min}=0.2\frac{R_{13}}{v_{s*,8.5}}\max\left[0.5,\frac{R_{13}^{0.4}}{(f_\rho\kappa_{0.34}M_{0})^{0.2}v_{s*,8.5}^{0.7}}\right]\,\rm days.
\end{equation}
The models assume a highly ionized envelope (RW11 deals with H, He and C/O envelopes while SW17 use a solar composition for the envelope), where the opacity is dominated by Thomson scattering, and a constant opacity approximation can be used. This approximation does not hold when the temperature drops below $\approx0.7\,\rm eV$ and a significant Hydrogen recombination takes place, which reduces the opacity sharply. The time at which the temperature is above $0.7\,\rm eV$ is given by
\begin{equation}\label{eq:t_opac}
t\lesssim t_{\rm opac}=6.47\left(\frac{v_{s*,8.5}^2}{f_{\rho}M_{0}\kappa_{0.34}}\right)^{0.061}\left(\frac{R_{13}}{\kappa_{0.34}}\right)^{0.56}\,\rm days.
\end{equation}
Even when the constant opacity approximation holds, RW11 breaks when the photosphere penetrates deeper into the progenitor shells and violates the model small $\delta$ approximation. The time at which the photosphere penetrates to a depth of $\delta\lesssim0.1$ is given by
\begin{equation}\label{eq:t_delta}
t<t_{\delta}=3f_\rho^{-0.1}\frac{\sqrt{\kappa_{0.34}M_{0}}}{v_{s*,8.5}}\,\rm days.
\end{equation}
RW11 results are independent of the progenitor density profile as the outer shells of the envelope, for which $\delta\ll1$, are universal for all progenitors with the same polytropic index $n$. SW17 extends RW11 for finite $\delta$, but introduces a dependency on the progenitor's density profile. SW17 is valid up to the point the envelope becomes almost transparent to radiation, $t_{\rm tr}/a$, where $t_{\rm tr}$ is the time the envelope is expected to become transparent and given by
\begin{equation}\label{eq:t_tr}
t_{\rm tr} \cong 19.5\left(\frac{\kappa_{0.34}M_{{\rm env},0}}{v_{s*,8.5}}\right)^{1/2}\,\rm days,
\end{equation}
$a$ is an order of unity parameter with the value 1.67 for $n=3/2$ and $M_{\rm env}=10^0M_{\rm env,0}M_\sun$ is the progenitor envelope mass. SW17 suppresses RW11 bolometric luminosity according to ($n=3/2$ values were used)
\begin{equation}\label{eq:SW17_L}
L/L_{\rm RW}=0.94\exp\left[-\left(\frac{at}{t_{\rm tr}}\right)^{0.8}\right].
\end{equation}
The SW17 validity extension is dependent on the envelope mass, $M_{\rm env}$, while the original RW11 part of the model depends on the ejecta mass $M_{\rm ej}$. These two masses are related by $M_{\rm ej} = M_{\rm env} + M_{\rm c}$, where $M_{\rm c}$ is the core mass \citep[this relation is described in SW17 \S2.2 and reviewed in][]{ArcaviEtAl2017}. The relation between $M_{\rm env}$ and $M_{\rm c}$ determines the value of $f_{\rho}$:
\begin{equation}\label{eq:f_rho}
f_{\rho}\approx\sqrt{M_{\rm env}/M_{\rm c}}.
\end{equation}
In addition to the validity extension, SW17 also recommend to use the ratio $f_T=T_{col}/T_{ph}$=1.1 rather than 1.2 suggested by RW11 due to the use of solar composition for the envelope instead of pure Hydrogen one.


\subsection{Morag-Sapir-Waxman model extension}
Morag et al. (2020, in preparation) extends the lower validity limit of the RW11 and SW17 models. It utilizes the asymptotic solution of the \citet[hereafter SKW13]{SapirEtAl2011,KatzEtAl2012,SapirEtAl2013} shock break out model as the solution for times earlier than $t_{\rm min}$ (Eq. \ref{eq:t_min}) and connect it to the SW17 shock cooling solution. The asymptotic solution of the break out model, translated to SW17 variables is given by
\begin{equation}\label{eq:L_SKW}
    L_{\rm SKW}=2.97\times10^{42}\frac{R_{13}^{2.46}v_{s*,8.5}^{0.60}}{(f_\rho M_0)^{0.06}\kappa_{0.34}^{1.06}}t_{\rm hr}^{-4/3}\,\rm erg\,s^{-1},
\end{equation}
where $t=1t_{\rm hr}\,\rm hr$, and
\begin{equation}\label{eq:T_SKW}
    T_{\rm SKW}=6.94\frac{R_{13}^{0.12}v_{s*,8.5}^{0.15}}{(f_\rho M_0)^{0.02}\kappa_{0.34}^{0.27}}t_{\rm hr}^{-1/3}\,\rm eV.
\end{equation}
The asymptotic luminosity of the break out solution decays $\propto t^{-4/3}$ while the SW17 luminosity is almost constant $\propto t^{-0.086}$. At $t_{\rm min}$ (Eq. \ref{eq:t_min}), when SW17 becomes valid, the breakout luminosity almost completely vanishes. Therefore the we can use the sum
\begin{equation}\label{eq:L_MSW}
    L_{\rm MSW}=L_{\rm SKW}+L_{\rm SW},
\end{equation}
to tie the two solutions. The asymptotic break out temperature decays slower ($\propto t^{-1/3}$) than the SW17 solution ($\propto t^{-1/2}$). Therefore, the transition between the temperatures of the planar and spherical solutions is given by
\begin{equation}\label{eq:T_MSW}
    T_{\rm MSW}=f_T \min[T_{\rm SKW},T_{\rm SW}].
\end{equation}
The full model which includes the SKW13 planar break out asymptotic solution and the SW17 spherical shock cooling solution was tested against numerical hydrodynamic simulations and was found to describe well (up to an error of +10\%/-30\%) the simulation results, starting at
\begin{equation}\label{eq:t_MSW}
    t_{\rm min,MSW}=3R_*/c=17R_{13}\,\rm min.
\end{equation}
At this time, the asymptotic break out solution is valid, and the spectrum is described well by a modified black body (i.e. with a shape of a black body spectrum, but with a luminosity different than $4\pi R^2\sigma_B T^4$), with $f_T=1.1$, and where $\sigma_B$ is the Stefan-Boltzmann constant. Although the SKW13 reference time is $t_{\rm BO}$, at $t_{\rm MSW}$ the photosphere is expanded enough and the difference between time elapsed since $t_{\rm BO}$ and the SW17 $t_{\rm ref}$ does not introduce a significant contribution to the bolometric luminosity or to the photosphere temperature, allowing both parts of the model to refer to $t_{\rm ref}$.

\subsection{{\normalfont{\texttt{SOPRANOS}}} Fitting Procedure}\label{subsec:SOPRANOS}
The challenge we face is to compare different models, where each model is valid for a different period of time and therefore for different subset of the data. The traditional fitting procedures compare all the models to a fixed subset of data points \citep[e.g.][]{ValentiEtAl2014,BoseEtAl2015,RubinEtAl2016,HosseinzadehEtAl2018}. The selection of a fixed subset of the data may limit the explored region of the parameters space and may prefer models whose validity region coincidentally matches the selected data subset over models with partial overlap. At an extreme case the method best fit may be invalidated by valid data points external to the fixed subset.

The \verb|SOPRANOS| (ShOck cooling modeling with saPiR \& wAxman model by gANOt \& Soumagnac) fitting procedure utilizes \textit{all} the valid data points corresponding to each model and \textit{only} those data points, by calculating the models likelihood. \verb|SOPRANOS| has two implementations. \verb|SOPRANOS-grid|, described in this paper and written in \verb|MATLAB|, which samples the parameter space by calculating the likelihood of a discrete grid of models, and \verb|SOPRANOS-nested|, described on \citet{SoumagnacEtAl2020} and written in python, which uses the nested sampling algorithm \citep{Skilling2004,Skilling2006,HigsonEtAl2019,Speagle2020} to dynamically sample the model parameter space.

Before fitting the model parameters with the \verb|SOPRANOS| algorithm we measure the SN redshifts from their spectra by identifying host galaxy lines. We also measure the $NUV$ background for each SN $NUV$ light curve from data points before and after the SN, in order to remove the host galaxy contribution from the SN measurements (both values appear in Table \ref{tab:SNtable}, the measurement of the $NUV$ background is described in \S\ref{subsec:background}). The redshift and $NUV$ background values are fixed and common for all the models of a SN and are not free model parameters we estimate.

With the redshift and background levels in hand, we calculate the expected flux in the two bands (\GALEX $NUV$ and PTF \textit{R}-band), for each model defined by a combination of the physical parameters using the MATLAB Astronomy and Astrophysics Toolbox \citep[MAAT;][]{Ofek2014}\footnote{\url{https://webhome.weizmann.ac.il/home/eofek/matlab/}} package. The bolometric luminosity, color temperature and the model validity period are calculated using the \verb|sn_cooling_msw| function. The spectrum of each band at the observer frame was calculated by
\begin{equation}\label{eq:spectrum}
    f_\lambda=\frac{L_{\rm MSW}}{4\pi D^2}\times\frac{(1+z)^4B_\lambda[T_{\rm col}/(1+z)]}{\sigma T_{\rm col}^4},
\end{equation}
where $L_{\rm MSW}$ is the photosphere bolometric luminosity, $T_{\rm col}=T_{\rm MSW}(t_d,f_T=1.1)$ is the color temperature (see Eq. \ref{eq:L_MSW}, \ref{eq:T_MSW}), $D$ is the luminosity distance and
\begin{equation}
    B_{\lambda}=\frac{2\pi h c^2}{\lambda^5}\frac{1}{\exp(\frac{hc}{\lambda k_B T})-1},
\end{equation}
is the Planck equation. $B_{\lambda,z}=(1+z)^4B_\lambda[T/(1+z)]$ is the redshifted Planck equation. The bolomotric flux $\int_0^\infty B_{\lambda,z}d\lambda=\int_0^\infty B_\lambda d\lambda = \sigma_B T^4$ is independent of redshift and therefore the temperature in the right hand side of the denominator of Eq. \ref{eq:spectrum} uses the color temperature at the SN frame.
With the predicted spectrum, and the filter transmission curves, we use the \verb|synphot| function for photon counting devices to calculate the exact flux expected for each of the bands. Note that \verb|synphot| applies an extinction correction for each bin of the input spectrum before integrating the spectrum to find the filter magnitude, as required. This method returns a different result compared with first calculating the filter magnitude and then applying the extinction using the filter pivot wavelength. This fact is important when dealing with the $NUV$ band, where the extinction curve has a significant slope.

To calculate $\mathcal{P}(\mathcal{M}_j|D)$, the posterior probability of the model $\mathcal{M}_j$, defined by the set of physical parameters \{$R_*^j$, $v_{s*}^j$, $M_{\rm ej}^j$, $f_{\rho}^j$, $t_{\rm ref}^j$, $E_{\rm B-V}^j$\}, where j is the model index, given the observation data set $D=\{t_{i}$,$f_i$,$\sigma_i\}$, we first find the subset of data points which are within the $\mathcal{M}_j$ valid period:
\begin{equation}
    \mathcal{V}_j=\left\{\{t_i,f_i,\sigma_i\}|t_{\rm MSW,z}^j\leq(t_i-t_{\rm ref}^j)\leq\min(t_{\rm opac,z}^j,t_{\rm tr,z}^j/a)\right\},
\end{equation}
where the index $i$ stands for the $i^{th}$ data point in $D$, and where $t^j_{\rm MSW,z}=(1+z)t^j_{\rm min,MSW}$, $t_{\rm opac,z}^j=(1+z)t_{\rm opac}^j$ and $t_{\rm tr,z}^j/a=(1+z)t_{\rm tr}^j/a$ are the redshifted $t_{\rm min,MSW}$, $t_{\rm opac}$ and $t_{\rm tr}$ for the model $\mathcal{M}_j$ (see Equations \ref{eq:t_opac}, \ref{eq:t_tr} and \ref{eq:t_MSW} and text above Equation \ref{eq:t_tr}). The model validity times are redshifted since the measurements were taken at the observer frame and not in the SN rest frame where those times are defined.

Assuming Gaussian statistics and that the data points are independent, we calculate the $\chi^2$ sum over the valid data points for the model $\mathcal{M}_j$
\begin{equation}
    \chi^2_j=\sum_{i\in \mathcal{V}_j}\frac{\{f_i-f^j[(t_i-t_{\rm ref}^j)/(1+z)]\}^2}{\sigma_{i}^2},
\end{equation}
where $f^j(t_d)$ is the predicted flux at time $t_d$ by the model $\mathcal{M}_j$, the time difference is blueshifted since the model is defined in the SN rest frame. The number of degrees of freedom for the model $\mathcal{M}_j$ is the number of valid data points ($N_j=|\mathcal{V}_j|$) less the number of free parameters we estimate, which is six for our case. We end this step by calculating the likelihood to measure those data points given the model $\mathcal{M}_j$:

\begin{equation}
    \mathcal{L}(D|\mathcal{M}_j)=\mathcal{P}_{\chi^2}(\chi_j^2,N_j-6),
\end{equation}
where $\mathcal{P}_{\chi^2}(\chi^2,N)$ is the $\chi^2$ probability distribution function with $N$ degrees of freedom. The
comparable quantity between models is their posterior probability, $\mathcal{P}(\mathcal{M}|D)$. To calculate it we utilize Bayes' theorem:
\begin{equation}
    \mathcal{P}(\mathcal{M}|D)=\frac{\mathcal{L}(D|\mathcal{M})\mathcal{P}(\mathcal{M})}{\mathcal{P}(D)},
\end{equation}
where $\mathcal{M}$ is the model, $D$ is the data, $\mathcal{P}(\mathcal{M})$ is the prior and $\mathcal{P}(D)$ is the probability to measure this data which is constant independent of the model. Now we can compare the posterior probabilities of different models. Since we are using a conservative flat prior (see \S\ref{subsec:prior}), and since $\mathcal{P}(D)$ is a constant independent of model, the posterior probability of a model is proportional to the likelihood to measure the data given the same model. We use the posterior probability to compare between models, and therefore we can ignore the constant proportion factor, and to compare the likelihood to measure our data given the different models as a discrimination measure between the different models.

At this point we have a grid of models, and for each model we have the likelihood to measure the data set $D=\{t_i,f_i,\sigma_i\}$ given this model, which is proportional to its posterior probability. Assuming the grid is dense enough, the shape of the discrete likelihood function describes well the behaviour of the continuous one. However, the values of the grid maxima are not the most likely models, since the grid models are limited to the arbitrary choice of the grid parameters values. To get the most likely models we numerically find the maxima of the continuous likelihood function, using the discrete grid maxima as initial conditions.

Recognizing $R_*$ and $v_{s*}$ are the dominant parameters of the models (see Eq. \ref{eq:RW11}, \ref{eq:L_SKW} and \ref{eq:T_SKW}), we start the optimization by marginalizing over $M_{\rm ej}$, $f_{\rho}$, $t_{\rm ref}$, and $E_{\rm B-V}$ and look for the local maxima of the two dimensional likelihood grid spanned by these two parameters. We ignore local maxima smaller than 1\% of the global grid maximum to avoid insignificant peaks resulting from numerical noise. For each one of the local maxima we repeat the following three stage optimization: At the first stage we allow only $R_*$ and $v_{s*}$ to vary, while for the rest of the parameters we use their grid values, and marginalizing over them to calculate the the likelihood. At the second stage we freeze $R_*$ and $v_{s*}$ values to the result of the first stage and allow $M_{\rm ej}$, $f_{\rho}$, $t_{\rm ref}$, and $E_{\rm B-V}$ to vary (their initial values are the maximum value of the four dimensional grid calculated in the first stage for the purpose of marginalization of those parameter). The last stage allows all the six parameters to vary, where the initial values are the result of the second stage. The numerical optimization uses MATLAB's \verb|fminsearch| function with $1-\mathcal{L}(D|\mathcal{M})$ as the objective function. We report the optimization results for all the local maxima in a table for each SN.

First results of the \verb|SOPRANOS| algorithm were already reported in \citet{SoumagnacEtAl2020} which used \verb|SOPRANOS-nested|. Here we use \verb|SOPRANOS-grid|. The two implementation are complimentary to each other. The grid implementation is useful to map the proper prior ranges for a SN (where the marginal likelihood distribution for each parameter converges asymptotically to 0, see \S\ref{subsec:prior}) and is independent of a specific sampling algorithm while the nested implementation has a higher resolution for the physical parameters and does not require a numerical optimization. Another advantage of these two implementation is the ability to test and verify them one against the other. The \verb|SOPRANOS-grid| implementation was added to the MAAT toolbox \citep{Ofek2014}.

\begin{table*}[ht]
\caption{GALEX/PTF SOPRANOS maximum likelihood models}
\centering
\begin{threeparttable}
\begin{tabular}{l l l l l l l l}
\hline
\hline
PTF name & $R_*$ & $v_{s*,8.5}$ & $M_{\rm ej}$ & $f_\rho$ & $t_{\rm ref}$ & $E_{\rm B-V}$          & $\chi^2/dof$ \\
\hline
PTF12gnt\tnote{a} & $418_{-2}^{+403}$ &$1.43_{-0.65}^{+0.00}$ & $2.2_{-0.2}^{+5.2}$  & $0.577_{-0.000}^{+0.575}$ & $56111.20_{-0.44}^{+0.00}$  & $0.052_{-0.006}^{+0.027}$ & 33.37/20\tnote{b}\\
PTF12gnt\tnote{a} & $636_{-216}^{+185}$ & $0.86_{-0.22}^{+0.17}$ & $3.0_{-1.0}^{+4.5}$  & $2.744_{-2.079}^{+0.027}$ & $56111.00_{-0.25}^{+0.19}$  & $0.048_{-0.001}^{+0.031}$ & 65.25/37\tnote{b}\\
PTF12ffs\tnote{c} & $788_{-244}^{+248}$ & $1.03_{-0.17}^{+0.21}$ & $5.0_{-3.0}^{+4.2}$ & $1.794_{-1.129}^{+0.192}$ & $56077.50_{-0.47}^{+0.54}$ & $0.040_{-0.015}^{+0.028}$ &  17.00/19\\
PTF12fhz &$ 249_{-149}^{+869}$ & $3.36_{-1.10}^{+0.00}$ & $ 2.5_{-2.0}^{+10.6}$ & $0.684_{-0.507}^{+0.000}$ &  $56086.06_{-0.33}^{+0.42}$  &  $0.089_{-0.002}^{+1.22}$ &   0.61/2          \\
PTF12fkp & $1262_{-511}^{+441}$ & $1.85_{-0.28}^{+0.27}$ &  $ 2.1_{- 0.1}^{+ 3.8}$  & $1.824_{-1.158}^{+0.101}$ & $56084.20_{-0.90}^{+0.14}$ &    $0.026_{-0.000}^{+0.061}$    &  12.90/14 \\
PTF12ftc & $ 895_{-621}^{+487}$ & $1.09_{-0.27}^{+0.43}$ &  $ 2.0_{- 0.0}^{+ 9.0}$ &  $2.395_{-1.628}^{+0.293}$ & $56089.49_{-1.15}^{+0.50}$ &       $0.038_{-0.001}^{+0.272}$  & 13.00/15 \\
PTF12glz\tnote{d} &             &                       &                         &                           &                             &                                 &          \\
\hline
\end{tabular}
\begin{tablenotes}
            \item[a] PTF12gnt likelihood map has multiple peaks. We report here two characterizing solutions.
            \item[b] PTF12gnt data include three suspicious points which do not fit any model, and contribute about 10 units to $\chi^2$.
            \item[c] PTF12ffs solutions have a degeneracy in $f_\rho$ and $M_{\rm ej}$ values, their multiplication is the measured quantity (see text).
            \item[d] PTF12glz is a type IIn SN which includes interaction with the CSM which is not modeled by MSW20. It is analyzed by \citet{SoumagnacEtAl2019a}.
\end{tablenotes}
\end{threeparttable}
\label{tab:results}
\end{table*}

\subsection{UV Background}\label{subsec:background}
The \verb|SOPRANOS| algorithm assumes the data points are background subtracted (including any host contribution) and all the measured flux comes from the SN. The \GALEX data analysis is not based on image subtraction and therefore may include some residual light from the SN host galaxy, and this contribution needs to be removed.

The data points before the SN rise are natural candidates to measure the $NUV$ background level. However, since some SNe were detected at the beginning of the experiment, we have only a small number of data points before the SN. The shock cooling models predict that the $NUV$ signal of the SN will decay and disappear as the ejecta cools down. We also see this behaviour in the observations themselves, which means that late $NUV$ data points after the $NUV$ transient may also be used for the measurement of the $NUV$ background.

The first step in measuring the background level is to identify the SN transient in the $NUV$. We compared all the data points to an initial reference background level from previous \GALEX images, where available, or to the lower $25^{th}$ percentile of data points, where previous archival \GALEX images were not available. We identify the SN transient beginning by a data point whose flux is larger than the reference background by more than $3\sigma$ and its end by the last data point with flux larger than the initial background by more than $1\sigma$. The different criteria for the beginning and the ending of the transient are a result of the model prediction of the photosphere cooling and therefore lower $NUV$ signal towards the end of the transient. If we were using a symmetrical criterion of $3\sigma$, data points which contain contribution from the SN would have been considered as not part of the transient, leading to an over estimation of the background level. We calculate the background value by comparing data points before and after the transient to a constant flux model, and minimize $\chi^2$ with respect to those points.

\subsection{Priors}\label{subsec:prior}
As described on section \ref{subsec:SOPRANOS}, we are fitting six parameters for the model $R_*$, $v_{s*}$, $M_{\rm ej}$, $f_{\rho}$, $t_{\rm ref}$ and $E_{\rm B-V}$ and using Bayes theorem. This requires some priors, which have to be selected carefully in order not to bias the results.  \citet{DaviesEtAl2018} explored the population of Red Super Giants in the Small and Large Maglanic Clouds and derived their radius out of the measured luminosity and temperatures \citep[See Figure 10 on][]{SoumagnacEtAl2020}. Most of the RSGs have radii smaller than 1100 solar radii, and only a few have radii in the range of 1100--1400 solar radii. We have chosen the range of 200--2000 solar radii as a flat conservative prior for the progenitor radius.

The $v_{s*}$ prior was selected iteratively, starting from a range of $1.6\times10^{8}-9.5\times10^{8}$ $\rm cm\,s^{-1}$ ($v_{s*,8.5}$=0.5-3) and adjusting it according to the marginalized distribution we receive until the tails of the marginalized distribution asymptotically reach 0. Typically this process converged to $v_{s*,8.5}$ values in the range of 0.4 to 5.
$M_{\rm ej}$ was chosen in the range of $2-20$ solar masses. $f_{\rho}$ was selected in the range of $\sqrt{1/3}$ and $\sqrt{10}$, reflecting the range of $M_{\rm c}/M_{\rm env}$ simulation cases tested by SW17 to validate their model (see Eq. \ref{eq:f_rho} for the relation between $f_{\rho}$ and $M_c/M_{\rm env}$). The $t_{\rm ref}$ prior was chosen from the last non-detection to the first detection in one of the two bands. $E_{\rm B-V}$ lower limit is constrained by the local galaxy extinction map of \citet{SchlegelEtAl1998} and the upper limit was selected iteratively according to the result marginal distribution, until its likelihood asymptotically converged to 0.

For PTF12fhz which has a double-peaked light curve, sometimes observed in other Type IIb SN, we chose some different priors. SW17 have shown their model may explain the first peak for $R_*\approx500R_\sun$ and $M_{\rm env}<1M_\sun$ (See their \S5).
We changed our priors for this SN to improve the resolution of the parameters in the regime described by SW17. The progenitor radius prior lower end was extended to $100R_\sun$, the ejecta mass lower limit was extended to $0.5M_\sun$ and its grid values was spanned logarithmically to improve the resolution for low masses. $f_\rho$ lower end was extended to $0.1$ to allow small values of $M_{\rm env}$. Note that SW17 was not tested against simulations with $f_\rho$ smaller than $\sqrt{1/3}$ and therefore the results for these low values should be taken with caution (SW17 used $f_\rho=0.3$ to describe this type of SNe).

\section{Results}\label{s:Results}

Table \ref{tab:results} summarizes the maximum likelihood model for each one of the GALEX/PTF SNe. For each SN we also present the spectra used to measure the redshift, the SN data points together with its maximum likelihood model-predicted light curves, the best-fit residuals for the maximum likelihood models, the marginalized $R_*$-$v_{s*,8.5}$ likelihood map, the marginalized distributions for $M_{\rm ej}$, $f_\rho$, $t_{\rm ref}$ and $E_{\rm B-V}$ and list all the models the \verb|SOPRANOS| algorithm converged to.

\subsection{PTF12ffs}
\begin{figure}[t]
    \centering
    \includegraphics[width=\linewidth]{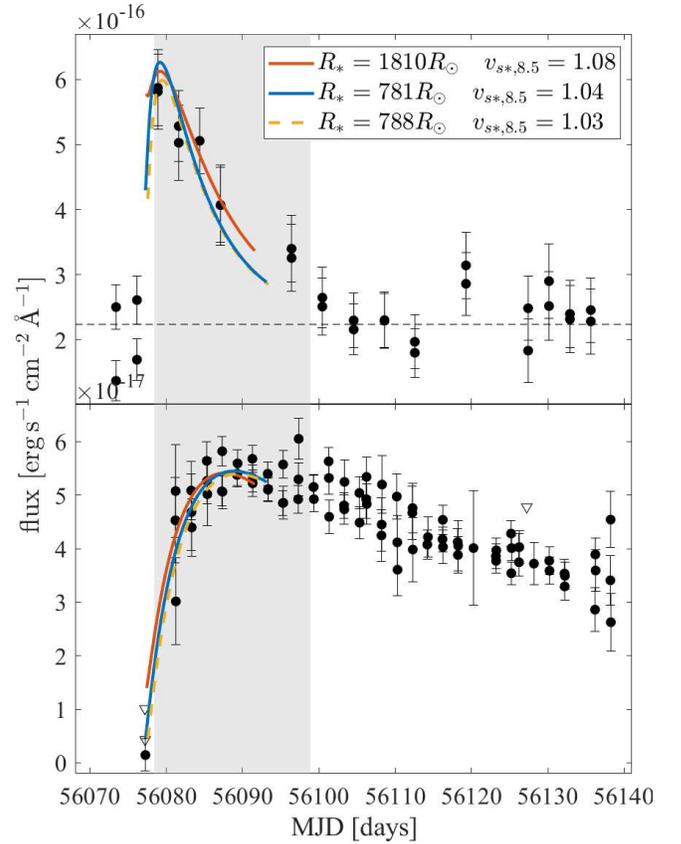}
    \caption{PTF12ffs data points and best fitting models. The upper panel describes the $NUV$ band data and the lower one is for the $R$ band. The dashed horizontal line in the upper panel represents the measured $NUV$ background. The triangles in the bottom panel stand for $3\sigma$ limits. The colored lines show the different solutions. The $R_*$ and $v_{s*}$ values for these solutions are listed in the legend while the other parameters are in Tables \ref{tab:ffs_results} and \ref{tab:ffs_results_Ebv0.09}. The grayed background area marks the $NUV$ transient. $NUV$ data points external to this area were used to calculate the $NUV$ background.}
    \label{fig:12ffs_LC}
\end{figure}

\begin{figure}[t]
    \centering
    \includegraphics[width=\linewidth]{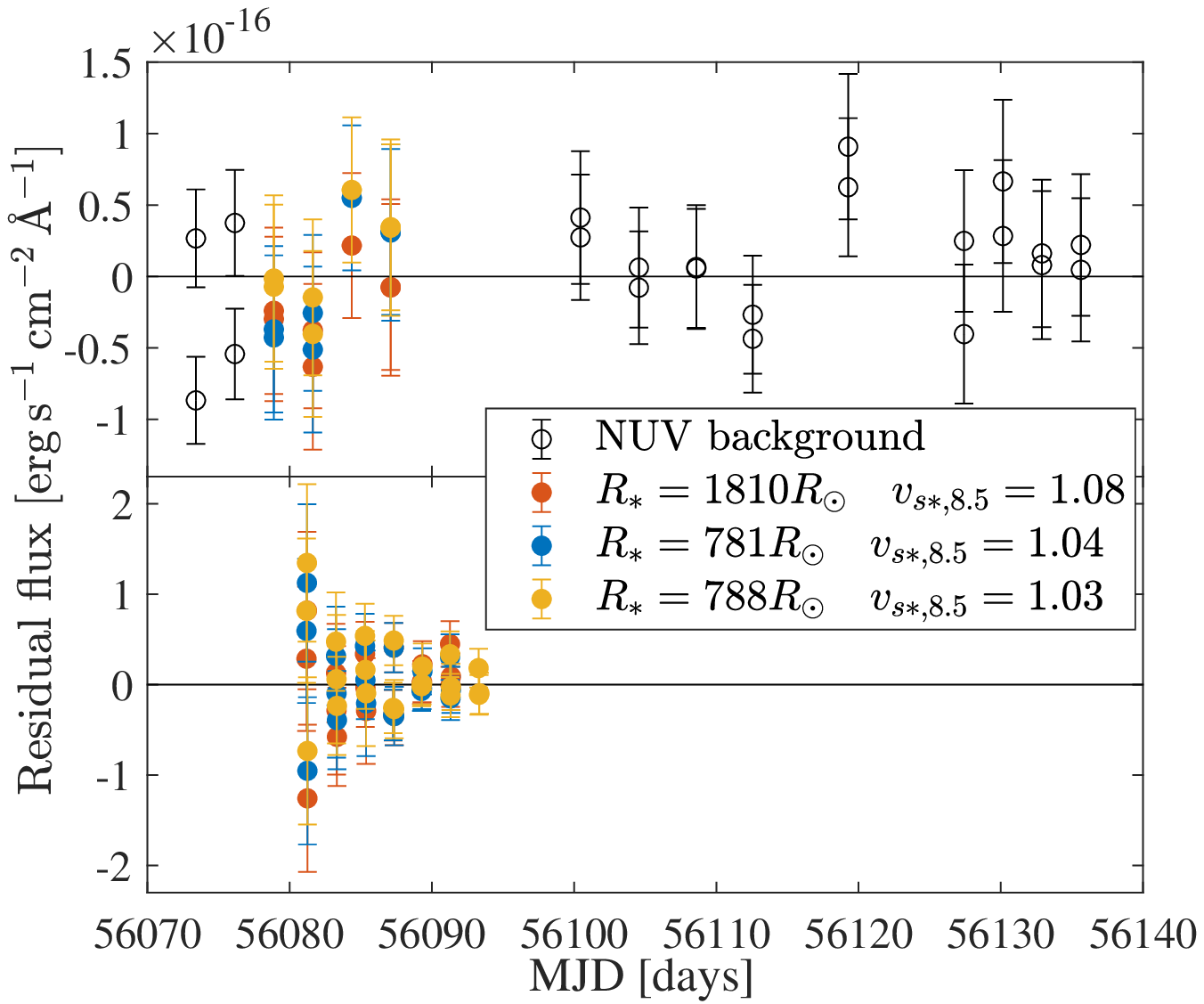}
    \caption{PTF12ffs best-fit residuals for the models plotted in Fig. \ref{fig:12ffs_LC}. The upper panel shows the $NUV$ band residuals and the bottom panel the $R$ band. The color code is the same as in Fig. \ref{fig:12ffs_LC}. The empty black circles are the residuals of the $NUV$ background estimation.}
    \label{fig:12ffs_Residuals}
\end{figure}

\begin{figure}
    \centering
    \includegraphics[width=\linewidth]{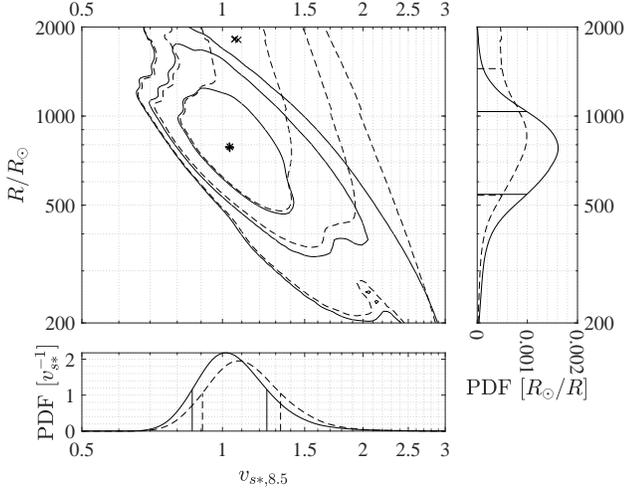}
    \caption{PTF12ffs $R_*$-$v_{s*}$ likelihood map. The solid contour lines represents 1$\sigma$, 2$\sigma$ and 3$\sigma$ of the cumulative likelihood with an extinction prior of $0.025\leq E_{B-V}\leq0.09$ Mag, and the dashed contour represents the cumulative likelihood with extinction prior of $0.025\leq E_{B-V}\leq0.26$ Mag. The plus markers indicate maximal likelihood models for the first extinction prior while the cross symbols are for the maximal likelihood models for the second extinction prior. The bottom and right panels show the marginal distribution for $v_{s*}$ and $R_*$ where the solid line plots the marginal distribution for the first prior and the dashed ones for the second. The vertical (horizontal) lines stand for the marginal distributions 1$\sigma$.}
    \label{fig:12ffs_CDF}
\end{figure}

\begin{figure}
    \centering
    \includegraphics[width=\linewidth]{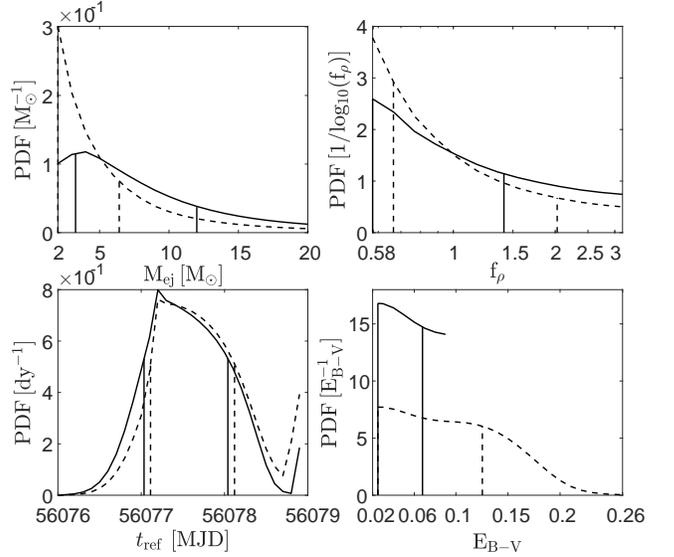}
    \caption{PTF12ffs Marginal distributions for $M_{\rm ej}$, $f_\rho$, $t_{\rm ref}$ and $E_{\rm B-V}$. Solid lines represent the marginal distributions for an extinction prior of up to 0.09 Mag while the dashed ones plot the distributions for a prior of up to 0.26 Mag.}
    \label{fig:12ffs_Marginals}
\end{figure}

\begin{table}[t]
\caption{PTF12ffs Likelihood local maxima \\$0.025\leq E_{B-V}\leq0.26$ Mag}
\centering
\begin{threeparttable}
\begin{tabular*}{\linewidth}{l l l l}
\hline
\hline
Parameter           & Peak \#1                    & Peak \#2                    & Peak \#3                    \\
\hline
$R_{*}$ [$R_\sun$]  & $1810_{-1042}^{+  0}$       & $1824_{-1049}^{+  0}$       & $ 786_{-246}^{+661}$        \\
$v_{s*,8.5}$        & $1.08_{-0.17}^{+0.25}$      & $1.06_{-0.16}^{+0.27}$      & $1.04_{-0.13}^{+0.29}$      \\
$M_{\rm ej}$ [$M_\sun$]& $ 2.0_{- 0.0}^{+ 4.4}$   & $ 2.0_{- 0.0}^{+ 4.4}$      & $12.0_{- 9.0}^{+ 1.3}$      \\
$t_{\rm ref}$ [MJD] & $56077.27_{-0.16}^{+0.85}$  & $56077.10_{-0.00}^{+1.01}$  & $56077.30_{-0.19}^{+0.82}$  \\
$f_{\rho}$          & $1.536_{-0.871}^{+0.491}$   & $1.556_{-0.890}^{+0.471}$   & $0.665_{-0.088}^{+0.523}$   \\
$E_{\rm B-V}$       & $0.133_{-0.102}^{+0.000}$   & $0.131_{-0.101}^{+0.000}$   & $0.030_{-0.005}^{+0.095}$   \\
$\chi^2$/dof        &  14.00/ 16                  &  14.23/ 16                  &  17.00/ 19                  \\
\hline
\hline
                    & Peak \#4                    & Peak \#5                    \\
\hline
$R_{*}$ [$R_\sun$]  & $ 789_{-249}^{+658}$        & $1817_{-1045}^{+  0}$       \\
$v_{s*,8.5}$        & $1.04_{-0.13}^{+0.30}$      & $1.06_{-0.15}^{+0.27}$      \\
$M_{\rm ej}$ [$M_\sun$]& $ 5.0_{- 3.0}^{+ 1.4}$   & $ 7.0_{- 4.0}^{+ 6.4}$      \\
$t_{\rm ref}$ [MJD] & $56077.60_{-0.49}^{+0.52}$  & $56077.40_{-0.29}^{+0.72}$  \\
$f_{\rho}$          & $1.794_{-1.129}^{+0.233}$   & $0.577_{-0.000}^{+0.611}$   \\
$E_{\rm B-V}$       & $0.040_{-0.015}^{+0.085}$   & $0.150_{-0.107}^{+0.000}$   \\
$\chi^2$/dof        &  17.00/ 19                  &  17.00/ 19                  \\
\hline
\end{tabular*}
\end{threeparttable}
\label{tab:ffs_results}
\end{table}

\begin{table}[t]
\caption{PTF12ffs Likelihood local maxima \\$0.025\leq E_{B-V}\leq0.09$ Mag}
\centering
\begin{threeparttable}
\begin{tabular}{l l l }
\hline
Parameter           & Peak \#1                    & Peak \#2                    \\
\hline
$R_{*}$ [$R_\sun$]  & $ 781_{-237}^{+254}$        & $ 788_{-244}^{+248}$        \\
$v_{s*,8.5}$        & $1.04_{-0.18}^{+0.21}$      & $1.03_{-0.17}^{+0.21}$      \\
$M_{\rm ej}$ [$M_\sun$]& $12.0_{- 8.7}^{+ 0.0}$   & $ 5.0_{- 3.0}^{+ 4.2}$      \\
$t_{\rm ref}$ [MJD] & $56077.20_{-0.17}^{+0.84}$  & $56077.50_{-0.47}^{+0.54}$  \\
$f_{\rho}$          & $0.665_{-0.088}^{+0.746}$   & $1.794_{-1.129}^{+0.192}$   \\
$E_{\rm B-V}$       & $0.030_{-0.005}^{+0.038}$   & $0.040_{-0.015}^{+0.028}$   \\
$\chi^2$/dof        &  17.00/ 19                  &  17.00/ 19                  \\
\hline
\hline
                    & Peak \#3                    & Peak \#4                    \\
\hline
$R_{*}$ [$R_\sun$]  & $ 782_{-238}^{+253}$        & $ 786_{-242}^{+249}$        \\
$v_{s*,8.5}$        & $1.04_{-0.18}^{+0.21}$      & $1.03_{-0.17}^{+0.21}$      \\
$M_{\rm ej}$ [$M_\sun$]& $ 5.0_{- 3.0}^{+ 4.2}$   & $12.0_{- 8.7}^{+ 0.0}$      \\
$t_{\rm ref}$ [MJD] & $56077.50_{-0.47}^{+0.54}$  & $56077.20_{-0.17}^{+0.84}$  \\
$f_{\rho}$          & $1.794_{-1.129}^{+0.192}$   & $0.665_{-0.088}^{+0.746}$   \\
$E_{\rm B-V}$       & $0.040_{-0.015}^{+0.028}$   & $0.030_{-0.005}^{+0.038}$   \\
$\chi^2$/dof        &  17.00/ 19                  &  17.00/ 19                  \\
\hline
\end{tabular}
\end{threeparttable}
\label{tab:ffs_results_Ebv0.09}
\end{table}

PTF12ffs data points together with its most likely models appear in Figure \ref{fig:12ffs_LC}, while the most likely best-fit residuals are presented in Figure \ref{fig:12ffs_Residuals}. The marginalized $R_*$-$v_{s*}$ likelihood map and the marginalized likelihood for each parameter are plotted in Fig. \ref{fig:12ffs_CDF} and the marginalized likelihood distributions for $M_{\rm ej}$, $f_\rho$, $t_{\rm ref}$ and $E_{\rm B-V}$ are shown in Fig. \ref{fig:12ffs_Marginals}. The local maxima of the likelihood function are listed in Table \ref{tab:ffs_results}.

Examining Figure \ref{fig:12ffs_CDF} we notice the likelihood distribution (dashed contour) includes two peaks (cross markers), one at $R_*\cong800R_\sun$ and the second at $R_*\cong1800R_\sun$. The radius marginal likelihood distribution has a non-negligible values for radius values greater than $2000R_*$, values which are not physical for the expected RSG progenitors. We try to separate between the two peaks in order to focus on the more physical solutions. The higher radius solutions are associated with higher extinction values of $E_{\rm B-V}\gtrsim0.13$ Mag while the lower radius solutions have lower extinction values of $E_{\rm B-V}\lesssim0.04$ Mag (Table \ref{tab:ffs_results}). This degeneracy is a result of the relation $L\propto R_*v_{s*}^2$ (Eq. \ref{eq:RW11}) which causes the higher radius solution to be more luminous. The extinction marginal distribution which appears in Fig. \ref{fig:12ffs_Marginals} (dashed line) has two local maxima. It includes a main peak at $E_{\rm B-V}\approx0.025$ Mag and a secondary peak at $E_{\rm B-V}\approx0.13$ Mag. We chose the point of $E_{B-V}=0.09$ Mag, where the extinction marginal likelihood distributions becomes flat, as the point which separates between the two peaks and repeated the analysis with narrower extinction prior which includes only values of $0.025\leq E_{\rm B-V}\leq0.09$ Mag. The results of the second analysis are plotted with solid lines on Figs. \ref{fig:12ffs_CDF} and \ref{fig:12ffs_Marginals}, and its maximal likelihood solutions are marked by plus symbols in Figure \ref{fig:12ffs_CDF} and are listed in Table \ref{tab:ffs_results_Ebv0.09}. Narrowing the extinction prior made the higher radius peak disappear and all the results models converged to a progenitor radius of $R_*\simeq790R_\sun$. The most likely models in Table \ref{tab:ffs_results_Ebv0.09} share similar values for all their parameters except for $f_\rho$ and $M_{\rm ej}$ which have a large scatter compared to the other parameters, and are separated into two subgroups: $f_\rho=0.665,\,M_{\rm ej}=12M_\sun$ and $f_\rho=1.794,\,M_{\rm ej}=5M_\sun$. While the two last parameter values have a large scatter, their multiplication, $8M_\sun$ and $9M_\sun$ respectively, has a small scatter, which is comparable to the one of the other parameters. Most of the model equations (\ref{eq:RW11}, \ref{eq:t_min}, \ref{eq:t_opac}) depend on the multiplication of the two. This degeneracy is removed only at late times by equations \ref{eq:t_tr} and \ref{eq:SW17_L}, in the case the photosphere becomes transparent before the recombination takes place. For our models $t_{\rm opac}\approx15$ days and $t_{\rm tr}\approx37$ days so recombination occurs before the photosphere becomes transparent, and the effects which depend on $M_{\rm env}$ become significant. Therefore we cannot solve the degeneracy between $f_\rho$ and $M_{\rm ej}$ for this SN.

\subsection{PTF12gnt}

\begin{figure}[t]
    \centering
    \includegraphics[width=\linewidth]{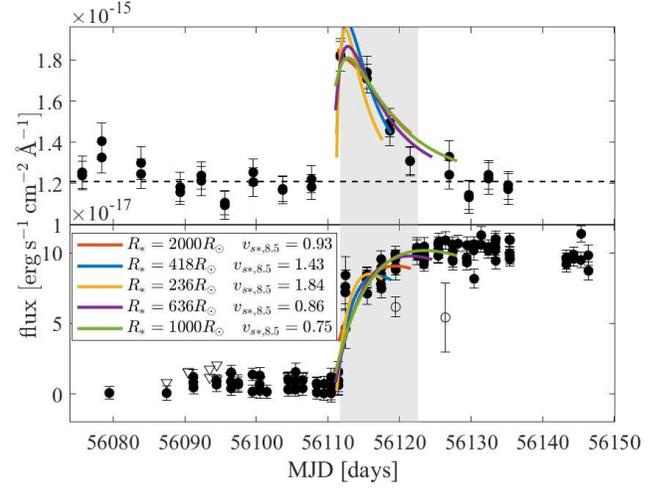}
    \caption{PTF12gnt data points and the best-fitting models. The upper panel describes the $NUV$ band data and the lower one is for the $R$ band. The dashed horizontal line in the top panel represents the measured $NUV$ background. The triangles in the bottom panel stand for $3\sigma$ limits and the empty circle error bars are outliers that were removed from the fit. The colored lines show the different solutions. The $R_*$ and $v_{s*}$ values of these solutions are listed in the legend while the other parameters are in Tables \ref{tab:gnt_results} and \ref{tab:gnt_results_Ebv0.15}. The gray background area marks the $NUV$ transient. Data points external to this area were used to calculate the $NUV$ background.}
    \label{fig:12gnt_LC}
\end{figure}

\begin{figure}[h]
    \centering
    \includegraphics[width=\linewidth]{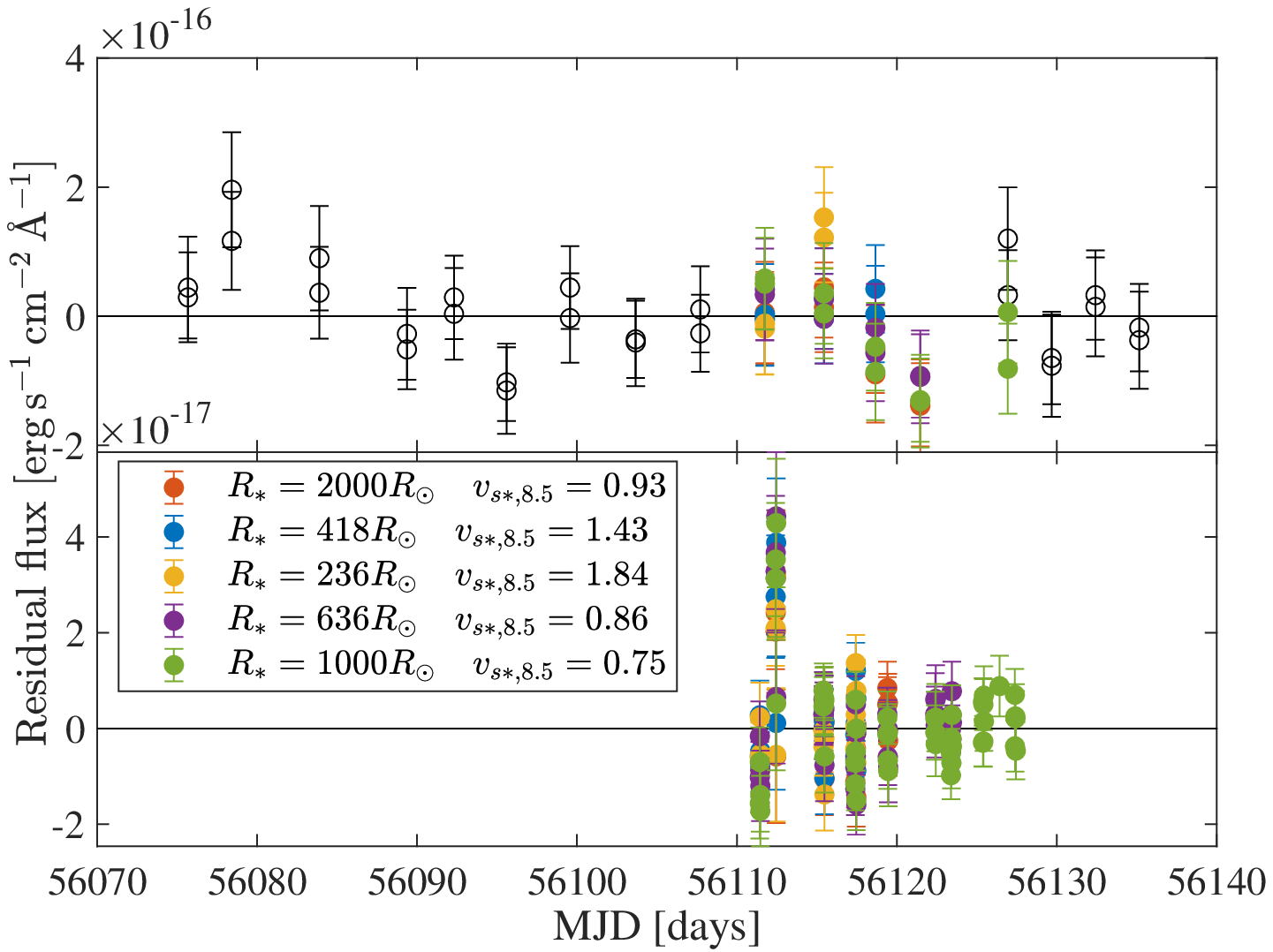}
    \caption{PTF12gnt best-fit residuals for the models plotted in Fig. \ref{fig:12gnt_LC}. The upper panel shows the $NUV$ band residuals and the bottom panel is for the $R$ band ones. The color code is the same as in Fig. \ref{fig:12ffs_LC}. The empty black circles are the residuals of the $NUV$ background estimation.}
    \label{fig:12gnt_Residuals}
\end{figure}

\begin{figure}
    \centering
    \includegraphics[width=\linewidth]{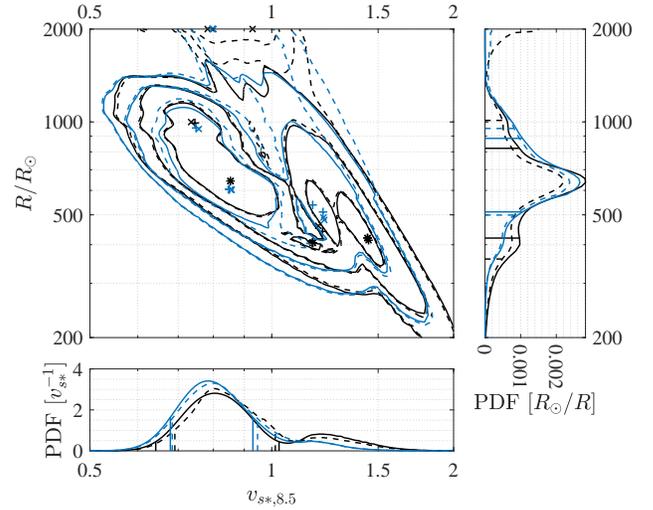}
    \caption{PTF12gnt $R_*$-$v_{s*}$ likelihood map. The solid contour lines represents 1$\sigma$, 2$\sigma$ and 3$\sigma$ of the cumulative likelihood with an extinction prior of $0.047\leq E_{B-V}\leq0.15$ Mag, and the dashed contour represents the cumulative likelihood with a prior of $0.047\leq E_{B-V}\leq0.26$ Mag. The plus markers indicate a maximal likelihood models with the first prior while the cross symbols are for the maximal likelihood models with the second prior (Asterisks are just plus and cross markers on the same spot and not a separate marker). The bottom and right panels show the marginal distributions for $v_{s*}$ and $R_*$ where the solid line plot the marginal distribution with the first prior and the dashed ones with the second prior. The black lines and symbols are the likelihood distributions and results when all the $R$ band data points were taken into account, and the blue ones represent the results when three suspicious $R$ band data points were treated as outliers.}
    \label{fig:12gnt_CDF}
\end{figure}
\begin{figure}
    \centering
    \includegraphics[width=\linewidth]{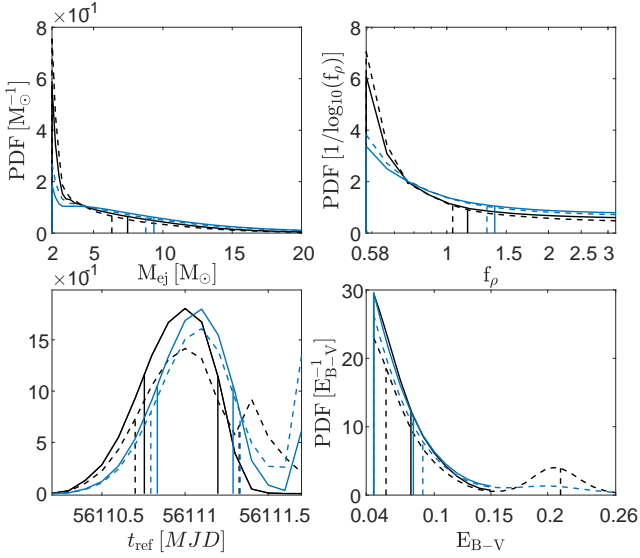}
    \caption{PTF12gnt Marginal distributions for $M_{\rm ej}$, $f_\rho$, $t_{\rm ref}$ and $E_{\rm B-V}$. Solid lines represent the marginal distributions with an extinction prior of up to 0.15 Mag while the dashed ones plot the distributions with a prior of up to 0.26 Mag. The black lines stand for the likelihood distributions when all the $R$ band data points were taken into account, and the blue ones represent the results when three suspicious $R$ band data points were treated as outliers.}
    \label{fig:12gnt_Marginals}
\end{figure}

\begin{table*}[t]
\caption{PTF12gnt Likelihood local maxima $0.046\leq E_{B-V}\leq0.26$ Mag}
\centering
\begin{threeparttable}
\begin{tabular*}{0.98\textwidth}{l l l l l l l l}
\hline
\hline
Parameter          & Peak \#1                    & Peak \#2                    & Peak \#3                    & Peak \#4
                   & Peak \#5                    & Peak \#6                    & Peak \#7                    \\
\hline
$R_{*}$ [$R_\sun$] & $2000_{-1372}^{+  0}$       & $ 414_{- 54}^{+598}$        & $ 421_{- 61}^{+591}$        & $ 420_{- 60}^{+592}$
                   & $ 460_{-101}^{+551}$        & $2000_{-1372}^{+  0}$       & $ 406_{- 46}^{+606}$        \\
$v_{s*,8.5}$       & $0.93_{-0.24}^{+0.08}$      & $1.45_{-0.63}^{+0.16}$      & $1.44_{-0.63}^{+0.16}$      & $1.44_{-0.63}^{+0.16}$
                   & $ 1.19_{-0.44}^{+0.00}$     & $0.78_{-0.09}^{+0.23}$      & $1.17_{-0.42}^{+0.00}$      \\
$M_{\rm ej}$ [$M_\sun$]&$ 2.6_{- 0.6}^{+ 3.8}$   & $ 2.2_{- 0.2}^{+ 4.1}$      & $ 2.2_{- 0.2}^{+ 4.1}$      & $ 2.2_{- 0.2}^{+ 4.1}$
                   & $ 4.2_{- 2.2}^{+ 2.1}$      & $ 2.0_{- 0.0}^{+ 4.3}$      & $ 2.3_{- 0.3}^{+ 4.1}$      \\
$t_{\rm ref}$ [MJD]& $56111.30_{-0.60}^{+0.03}$  & $56111.20_{-0.50}^{+0.13}$  & $56111.19_{-0.49}^{+0.13}$  & $56111.20_{-0.50}^{+0.13}$
                   & $56111.10_{-0.40}^{+0.23}$  & $56111.31_{-0.61}^{+0.02}$  & $56111.03_{-0.33}^{+0.29}$  \\
$f_{\rho}$         & $0.577_{-0.000}^{+0.464}$   & $0.577_{-0.000}^{+0.464}$   & $0.579_{-0.001}^{+0.462}$   & $0.577_{-0.000}^{+0.464}$
                   & $0.577_{-0.000}^{+0.464}$   & $1.840_{-1.176}^{+1.323}$   & $1.338_{-0.674}^{+1.825}$   \\
$E_{\rm B-V}$      & $0.208_{-0.151}^{+0.003}$   & $0.053_{-0.007}^{+0.048}$   & $0.054_{-0.008}^{+0.047}$   & $0.055_{-0.008}^{+0.047}$
                   & $0.050_{-0.004}^{+0.051}$   & $0.209_{-0.151}^{+0.003}$   & $0.047_{-0.000}^{+0.055}$   \\
$\chi^2$/dof       &  33.63/ 21                  &  33.24/ 20                  &  33.28/ 20                  &  33.28/ 20
                   &  49.44/ 27                 &  75.11/ 47                  &  51.28/ 27                   \\
\hline
\hline
                   & Peak \#8                    & Peak \#9                    & Peak \#10                   & Peak \#11
                   & Peak \#12                   & Peak \#13                    \\
\hline
$R_{*}$ [$R_\sun$] & $ 405_{- 46}^{+606}$        & $ 406_{- 47}^{+605}$        & $ 406_{- 46}^{+606}$       & $ 643_{-284}^{+368}$
                   & $ 999_{-640}^{+ 13}$        & $ 643_{-284}^{+368}$        \\
$v_{s*,8.5}$       & $1.17_{-0.42}^{+0.00}$      & $1.17_{-0.42}^{+0.00}$      & $1.17_{-0.42}^{+0.00}$     & $0.85_{-0.16}^{+0.16}$
                   & $0.74_{-0.04}^{+0.28}$      & $0.85_{-0.16}^{+0.16}$      \\
$M_{\rm ej}$ [$M_\sun$]&$ 2.3_{- 0.3}^{+ 4.1}$   & $ 4.1_{- 2.1}^{+ 2.2}$      & $ 2.2_{- 0.2}^{+ 4.1}$     & $ 2.8_{- 0.8}^{+ 3.5}$
                   & $ 3.5_{- 1.5}^{+ 2.8}$      & $ 2.5_{- 0.5}^{+ 3.8}$      \\
$t_{\rm ref}$ [MJD]& $56111.04_{-0.34}^{+0.28}$  & $56111.03_{-0.33}^{+0.29}$  & $56111.04_{-0.34}^{+0.29}$ & $56110.97_{-0.27}^{+0.35}$
                   & $56110.96_{-0.26}^{+0.37}$  & $56111.30_{-0.60}^{+0.03}$  \\
$f_{\rho}$         & $1.342_{-0.678}^{+1.821}$   & $0.740_{-0.162}^{+0.301}$   & $1.353_{-0.689}^{+1.809}$  & $3.162_{-2.498}^{+0.000}$
                   & $3.073_{-2.409}^{+0.090}$   & $0.577_{-0.000}^{+0.464}$   \\
$E_{\rm B-V}$      & $0.047_{-0.000}^{+0.055}$   & $0.047_{-0.000}^{+0.055}$   & $0.047_{-0.000}^{+0.055}$  & $0.049_{-0.002}^{+0.052}$
                   & $0.091_{-0.044}^{+0.011}$   & $0.210_{-0.152}^{+0.001}$   \\
$\chi^2$/dof       &  51.28/ 27                  &  51.29/ 27                  &  51.29/ 27                 &  65.23/ 37
                   &  88.85/ 51                  & 1418.86/ 27                 \\
\hline
\end{tabular*}
\end{threeparttable}
\label{tab:gnt_results}
\end{table*}

\begin{table*}[t]
\caption{PTF12gnt Likelihood local maxima $0.046\leq E_{B-V}\leq0.15$ Mag}
\centering
\begin{threeparttable}
\begin{tabular}{l l l l l l}
\hline
\hline
Parameter          & Peak \#1                    & Peak \#2                    & Peak \#3                    & Peak \#4                    & Peak \#5                  \\
\hline
$R_{*}$ [$R_\sun$] &   $ 418_{-2}^{+403}$        & $ 455_{- 34}^{+367}$        & $ 236_{-  0}^{+512}$        & $ 491_{- 71}^{+330}$        & $ 382_{-  0}^{+404}$      \\
$v_{s*,8.5}$       & $1.43_{-0.65}^{+0.00}$      & $1.39_{-0.62}^{+0.00}$      & $1.84_{-1.03}^{+0.16}$      & $1.21_{-0.48}^{+0.00}$      & $1.25_{-0.51}^{+0.00}$    \\
$M_{\rm ej}$ [$M_\sun$]&$ 2.2_{- 0.2}^{+ 5.2}$   & $ 2.2_{- 0.2}^{+ 5.2}$      & $ 2.0_{- 0.0}^{+ 5.5}$      & $ 3.7_{- 1.7}^{+ 3.8}$      & $ 2.8_{- 0.8}^{+ 4.7}$    \\
$t_{\rm ref}$ [MJD]& $56111.20_{-0.44}^{+0.00}$  & $56111.20_{-0.44}^{+0.00}$  & $56111.20_{-0.44}^{+0.00}$  & $56111.10_{-0.35}^{+0.09}$  & $56111.10_{-0.35}^{+0.09}$\\
$f_{\rho}$         & $0.577_{-0.000}^{+0.575}$   & $0.577_{-0.000}^{+0.575}$   & $0.577_{-0.000}^{+0.575}$   & $0.577_{-0.000}^{+0.575}$   & $0.883_{-0.306}^{+0.269}$ \\
$E_{\rm B-V}$      & $0.052_{-0.006}^{+0.027}$   & $0.055_{-0.008}^{+0.024}$   & $0.050_{-0.003}^{+0.029}$   & $0.055_{-0.008}^{+0.024}$   & $0.047_{-0.000}^{+0.033}$ \\
$\chi^2$/dof       &  33.37/ 20                  &  33.89/ 20                  &  27.62/ 14                  &  49.93/ 27                  &  50.92/ 27                \\
\hline
\hline
                   & Peak \#6                    & Peak \#7                    & Peak \#8                    & Peak \#9                    & Peak \#10                 \\
\hline
$R_{*}$ [$R_\sun$] & $ 400_{-  0}^{+402}$        & $ 636_{-216}^{+185}$        & $ 418_{ -2}^{+403}$        & $ 436_{- 16}^{+385}$        & $1000_{-457}^{+  0}$      \\
$v_{s*,8.5}$       & $1.16_{-0.45}^{+0.00}$      & $0.86_{-0.22}^{+0.17}$      & $1.12_{-0.43}^{+0.00}$      & $1.06_{-0.39}^{+0.00}$      & $0.75_{-0.10}^{+0.28}$    \\
$M_{\rm ej}$ [$M_\sun$]& $ 3.0_{- 1.0}^{+ 4.5}$  & $ 3.0_{- 1.0}^{+ 4.5}$      & $ 3.3_{- 1.3}^{+ 4.1}$      & $ 2.0_{- 0.0}^{+ 5.5}$      & $ 3.7_{- 1.7}^{+ 3.8}$    \\
$t_{\rm ref}$ [MJD]& $56111.00_{-0.25}^{+0.19}$  & $56111.00_{-0.25}^{+0.19}$  & $56111.00_{-0.25}^{+0.19}$  & $56111.00_{-0.25}^{+0.19}$  & $56110.90_{-0.15}^{+0.29}$\\
$f_{\rho}$         & $1.018_{-0.440}^{+0.135}$   & $2.744_{-2.079}^{+0.027}$   & $1.018_{-0.440}^{+0.135}$   & $2.744_{-2.079}^{+0.027}$   & $2.382_{-1.717}^{+0.389}$ \\
$E_{\rm B-V}$      & $0.047_{-0.000}^{+0.033}$   & $0.048_{-0.001}^{+0.031}$   & $0.047_{-0.000}^{+0.033}$   & $0.047_{-0.000}^{+0.033}$   & $0.090_{-0.037}^{+0.007}$ \\
$\chi^2$/dof       &  52.06/ 27                  &  65.25/ 37                  &  52.41/ 27                  &  53.21/ 27                  &  89.67/ 51                \\
\hline
\end{tabular}
\end{threeparttable}
\label{tab:gnt_results_Ebv0.15}
\end{table*}

\begin{figure}[h]
    \centering
    \includegraphics[width=\linewidth]{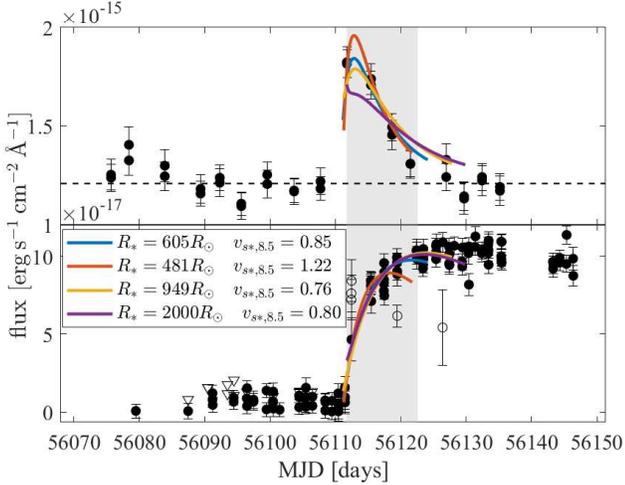}
    \caption{PTF12gnt data points and best-fitting models when the suspicious three $R$ band-data points were treated as outliers. The same symbol convention as in Fig \ref{fig:12gnt_LC} is used. The model full parameters are listed in Tables \ref{tab:gnt_ol_results} and \ref{tab:gnt_ol_results_Ebv0.15}.}
    \label{fig:12gnt_ol_LC}
\end{figure}

\begin{table*}
\caption{PTF12gnt Likelihood local maxima, without 3 $R$ data points, $0.046\leq E_{B-V}\leq0.26$ Mag}
\centering
\begin{threeparttable}
\begin{tabular}{l l l l l l}
\hline
\hline
Parameter           & Peak \#1                    & Peak \#2                    & Peak \#3                    & Peak \#4                    \\
\hline
$R_{*}$ [$R_\sun$]  & $ 481_{-   0}^{+ 455}$      & $ 490_{-   0}^{+ 454}$      & $ 604_{- 105}^{+ 349}$      & $ 603_{- 104}^{+ 350}$      \\
$v_{s*,8.5}$        & $1.22_{-0.46}^{+0.00}$      & $1.22_{-0.45}^{+0.00}$      & $0.85_{-0.17}^{+0.09}$      & $0.86_{-0.17}^{+0.09}$      \\
$M_{\rm ej}$ [$M_\sun$]& $ 3.4_{- 1.4}^{+ 5.4}$   & $ 3.4_{- 1.4}^{+ 5.4}$      & $ 3.1_{- 1.1}^{+ 5.6}$      & $ 3.1_{- 1.1}^{+ 5.7}$      \\
$t_{\rm ref}$ [MJD] & $56111.23_{-0.44}^{+0.10}$  & $56111.23_{-0.43}^{+0.10}$  & $56111.11_{-0.32}^{+0.22}$  & $56111.12_{-0.32}^{+0.21}$  \\
$f_{\rho}$          & $0.596_{-0.019}^{+0.719}$   & $0.594_{-0.017}^{+0.721}$   & $3.161_{-2.416}^{+0.002}$   & $3.161_{-2.416}^{+0.001}$   \\
$E_{\rm B-V}$       & $0.048_{-0.001}^{+0.042}$   & $0.051_{-0.005}^{+0.039}$   & $0.047_{-0.000}^{+0.043}$   & $0.047_{-0.000}^{+0.043}$   \\
$\chi^2$/dof        &  24.83/ 22                  &  24.87/ 22                  &  35.77/ 34                  &  35.82/ 34                  \\
\hline
\hline
                    & Peak \#5                    & Peak \#6                    & Peak \#7                    & Peak \#8                    \\
\hline
$R_{*}$ [$R_\sun$]  & $2000_{-1361}^{+   0}$      & $2000_{-1361}^{+   0}$      & $2000_{-1361}^{+   0}$      & $ 949_{- 450}^{+   3}$      \\
$v_{s*,8.5}$        & $0.80_{-0.12}^{+0.15}$      & $0.80_{-0.11}^{+0.15}$      & $0.80_{-0.11}^{+0.15}$      & $0.76_{-0.07}^{+0.19}$      \\
$M_{\rm ej}$ [$M_\sun$]& $ 4.0_{- 2.0}^{+ 4.8}$   & $ 4.0_{- 2.0}^{+ 4.8}$      & $ 2.4_{- 0.4}^{+ 6.3}$      & $14.2_{-10.6}^{+ 0.8}$      \\
$t_{\rm ref}$ [MJD] & $56111.58_{-0.68}^{+0.00}$  & $56111.58_{-0.68}^{+0.00}$  & $56111.59_{-0.69}^{+0.00}$  & $56111.10_{-0.31}^{+0.23}$  \\
$f_{\rho}$          & $0.890_{-0.312}^{+0.425}$   & $0.886_{-0.308}^{+0.429}$   & $1.602_{-0.970}^{+0.192}$   & $0.577_{-0.000}^{+0.738}$   \\
$E_{\rm B-V}$       & $0.232_{-0.172}^{+0.001}$   & $0.230_{-0.170}^{+0.000}$   & $0.233_{-0.173}^{+0.007}$   & $0.090_{-0.042}^{+0.005}$   \\
$\chi^2$/dof        &  55.03/ 45                  &  55.05/ 45                  &  55.05/ 45                  &  60.25/ 48                  \\
\hline
\end{tabular}
\end{threeparttable}
\label{tab:gnt_ol_results}
\end{table*}

\begin{table}
\caption{PTF12gnt Likelihood local maxima, without 3 $R$ data points, $0.046\leq E_{B-V}\leq0.15$ Mag}
\centering
\begin{threeparttable}
\begin{tabular}{l l l l}
\hline
\hline
Parameter          & Peak \#1                    & Peak \#2                    \\
\hline
$R_{*}$ [$R_\sun$] & $ 511_{-   0}^{+ 374}$      & $ 538_{-  27}^{+ 347}$      \\
$v_{s*,8.5}$       & $1.22_{-0.46}^{+0.00}$      & $1.17_{-0.42}^{+0.00}$      \\
$M_{\rm ej}$ [$M_\sun$]&$ 2.5_{- 0.5}^{+ 6.9}$   & $ 3.0_{- 1.0}^{+ 6.3}$      \\
$t_{\rm ref}$ [MJD]& $56111.23_{-0.40}^{+0.06}$  & $56111.18_{-0.35}^{+0.11}$  \\
$f_{\rho}$         & $0.736_{-0.158}^{+0.651}$   & $0.711_{-0.133}^{+0.676}$   \\
$E_{\rm B-V}$      & $0.053_{-0.006}^{+0.029}$   & $0.066_{-0.019}^{+0.016}$   \\
$\chi^2$/dof       &  25.04/ 22                  &  26.82/ 24                  \\
\hline
\hline
                    & Peak \#3                    & Peak \#4                    \\
\hline
$R_{*}$ [$R_\sun$] & $ 605_{-  94}^{+ 280}$      & $ 956_{- 396}^{+   0}$      \\
$v_{s*,8.5}$       & $0.85_{-0.17}^{+0.08}$      & $0.75_{-0.07}^{+0.18}$      \\
$M_{\rm ej}$ [$M_\sun$]& $ 3.2_{- 1.2}^{+ 6.1}$  & $14.6_{-10.7}^{+ 0.0}$      \\
$t_{\rm ref}$ [MJD]& $56111.11_{-0.28}^{+0.18}$  & $56111.09_{-0.25}^{+0.20}$  \\
$f_{\rho}$         & $3.162_{-2.397}^{+0.000}$   & $0.577_{-0.000}^{+0.809}$   \\
$E_{\rm B-V}$      & $0.047_{-0.000}^{+0.035}$   & $0.091_{-0.041}^{+0.001}$   \\
$\chi^2$/dof       &  35.72/ 34                  &  60.55/ 48                  \\
\hline
\end{tabular}
\end{threeparttable}
\label{tab:gnt_ol_results_Ebv0.15}
\end{table}

PTF12gnt data points together with its most likely models appear in Figure \ref{fig:12gnt_LC}. The residuals of the most likely best-fit are presented in Figure \ref{fig:12gnt_Residuals}. The marginalized $R_*$-$v_{s*}$ likelihood map and the marginalized likelihood for each parameter are plotted in Figure \ref{fig:12gnt_CDF} and the marginalized likelihood distributions for $M_{\rm ej}$, $f_\rho$, $t_{\rm ref}$ and $E_{\rm B-V}$ are shown in Figure \ref{fig:12gnt_Marginals}. The local maxima of the likelihood function are listed in Table \ref{tab:gnt_results} and marked on Figure \ref{fig:12gnt_Residuals} by cross symbols. As for PTF12ffs, looking at the likelihood distribution map in Fig. \ref{fig:12gnt_CDF} (dashed black contours), we can distinguish between solutions with progenitor radius greater and smaller than $1000R_\sun$ by splitting the extinction prior to values smaller or greater than $E_{\rm B-V}=0.15$ Mag. The extinction-marginalized likelihood distribution (dashed black line in Fig. \ref{fig:12gnt_Marginals}) includes two well separated peaks located at $E_{\rm B-V}=0.026$ Mag and $E_{\rm B-V}=0.21$ Mag, and $E_{\rm B-V}=0.15$ Mag is the minimum between them. When limiting the extinction-prior to $0.047\leq E_{\rm B-V}\leq0.15$ Mag the solutions with progenitor radius greater than $1000R_\sun$ disappear (solid black contour on Fig. \ref{fig:12gnt_CDF}) and the secondary peak in $t_{\rm ref}$ (dashed and solid black lines in Fig. \ref{fig:12gnt_Marginals}) disappears as well. The maximal likelihood solutions with narrowed extinction prior (Table \ref{tab:gnt_results_Ebv0.15}, plus black markers on Fig. \ref{fig:12gnt_CDF}) are not located in a single peak like PTF12ffs but include several sub-peaks. The left cluster of solutions is characterized by higher progenitor radii ($500-1000R_\sun$) and lower shock velocity parameter. The solutions in this cluster include a non negligible likelihood for all  the $f_\rho$ values. The right cluster of solutions is characterized by lower progenitor radii ($350-700R_*$) and higher shock velocity parameter values. The solutions in this cluster have a significant likelihood only for $f_\rho$ values smaller than 1.2. Although the likelihood of the individual solutions in the right cluster have a higher likelihood than the ones in the left cluster (see Table \ref{tab:gnt_results_Ebv0.15}), the marginalized distributions for $R_*$ and $v_{s*}$, which are a result of an integral over all of the $f_\rho$ values have a higher likelihood values than those of the left cluster solutions.

All of the solutions for the SN have a $\chi^2$/dof ratio of $1.5-1.8$ which is worse than the ratio we receive for the other SNe. Examining the light curve (Figure \ref{fig:12gnt_LC}) and the best-fit residuals (Fig. \ref{fig:12gnt_Residuals}) we see that the $R$-band measurements from the first night of the SN detection are spread along a large range of flux values. All the solutions (Tables \ref{tab:gnt_results} and \ref{tab:gnt_results_Ebv0.15}) agree with the lowest flux point measured and fail to match the higher flux values during this night. Those points introduce an error of about 10 in $\chi^2$ units for all the solutions, leading to this unusual goodness of fit ratio. Since the $\chi^2$ distribution has the highest slope when $\chi^2\sim \nu$, where $\nu$ is the number of degrees of freedom, the effect of those points on models with higher likelihood (i.e. $\chi^2\sim \nu$) is larger than their effect on models with lower likelihood. When we ignore these three measurements and treat them as outliers, we get better goodness of fit scores as presented in Tables \ref{tab:gnt_ol_results} and \ref{tab:gnt_ol_results_Ebv0.15}. Those scores are similar to the ones we get for the other SNe. Removing those points also changes the marginalized progenitor radius likelihood (Blue contours in Fig. \ref{fig:12gnt_CDF}) and suppresses the secondary peaks at $\sim2000R_\sun$ and $R_*\sim400R_\sun$. We still find a correlation between the $R_*\sim2000R_\sun$ solutions and the extinction values larger than $E_{\rm B-V}=0.15$ Mag and present the results with a narrower extinction prior. Selected solutions from Tables \ref{tab:gnt_ol_results} and \ref{tab:gnt_ol_results_Ebv0.15} are plotted against the data in Figure \ref{fig:12gnt_ol_LC}. We see that while all the solutions match the $R$-band data points well, the $R_*=2000R_\sun$ solution does not match the early $NUV$ data points.

\subsection{PTF12fkp}
\begin{figure}
    \centering
    \includegraphics[width=\linewidth]{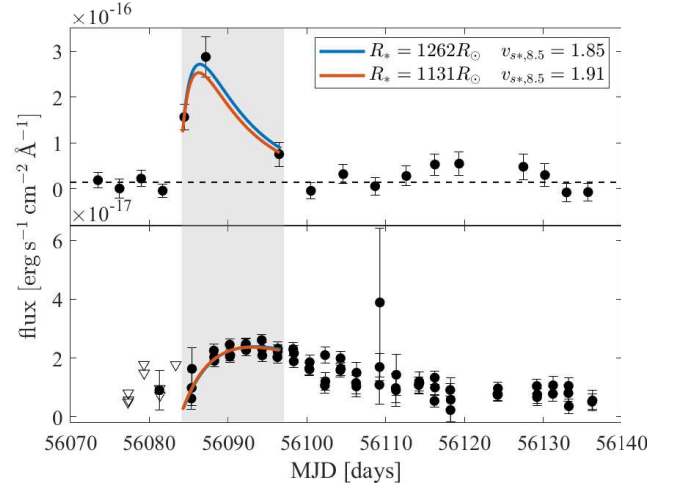}
    \caption{PTF12fkp data points and best-fitting models. The upper panel describes the $NUV$ band data and the lower one is for the $R$ band. The dashed horizontal line in the top panel represents the measured $NUV$ background. The triangles in the bottom panel stand for $3\sigma$ limits. The colored lines show the different solutions. The $R_*$ and $v_{s*}$ values of these solutions are list in the legend, while the other parameters are listed in Table \ref{tab:fkp_results}. The grayed background area indicates the $NUV$ transient. $NUV$ data points external to this area were used to calculate the $NUV$ background.}
    \label{fig:12fkp_LC}
\end{figure}

\begin{figure}
    \centering
    \includegraphics[width=\linewidth]{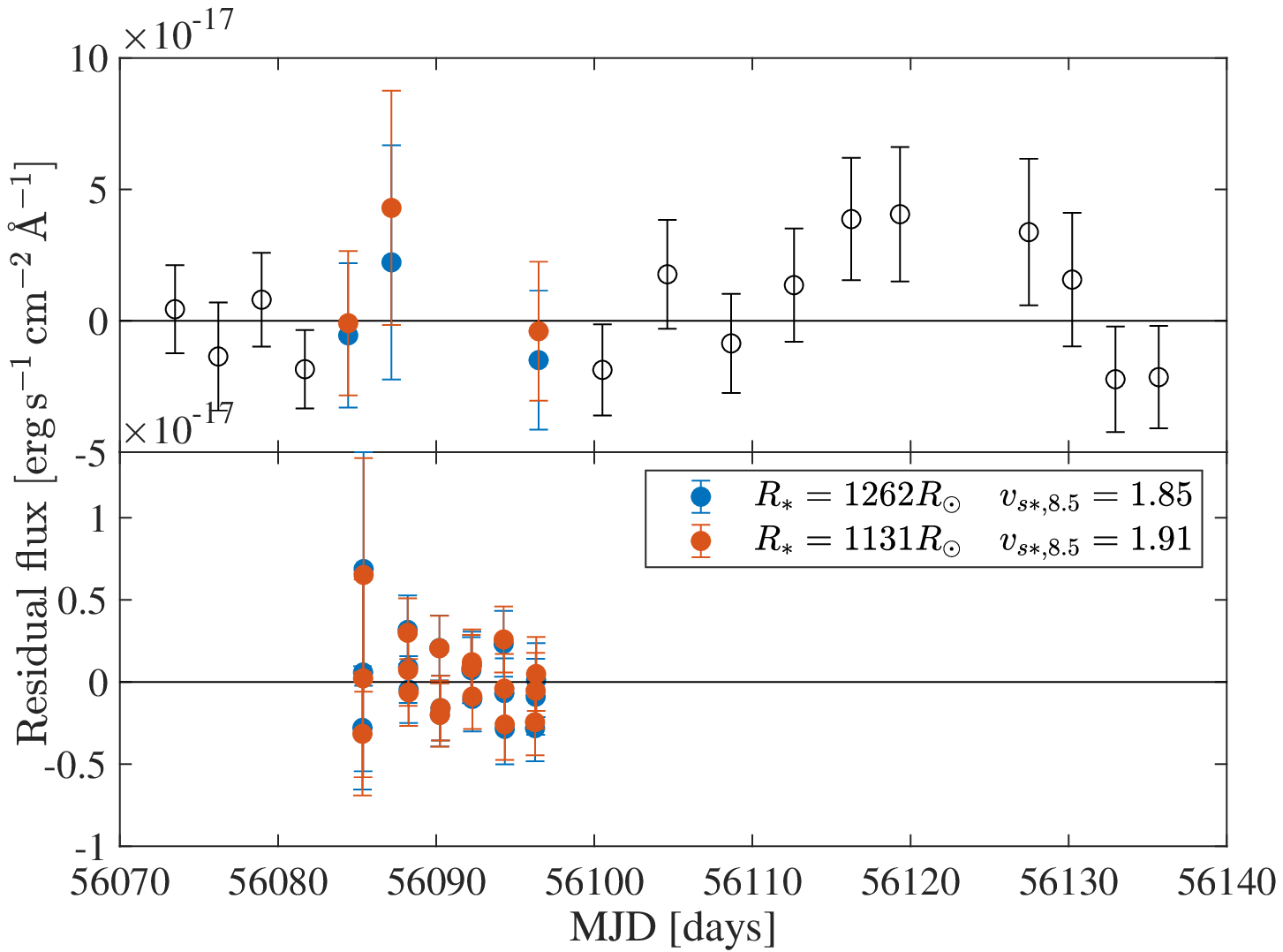}
    \caption{PTF12fkp best-fit residuals for the models plotted in Fig. \ref{fig:12fkp_LC}. The upper panel shows the $NUV$ band residuals and the lower panel the $R$ band ones. The color code is the same as in Fig. \ref{fig:12fkp_LC}. The empty black circles are the residuals of the $NUV$ background estimation.}
    \label{fig:12fkp_Residuals}
\end{figure}

\begin{figure}
    \centering
    \includegraphics[width=\linewidth]{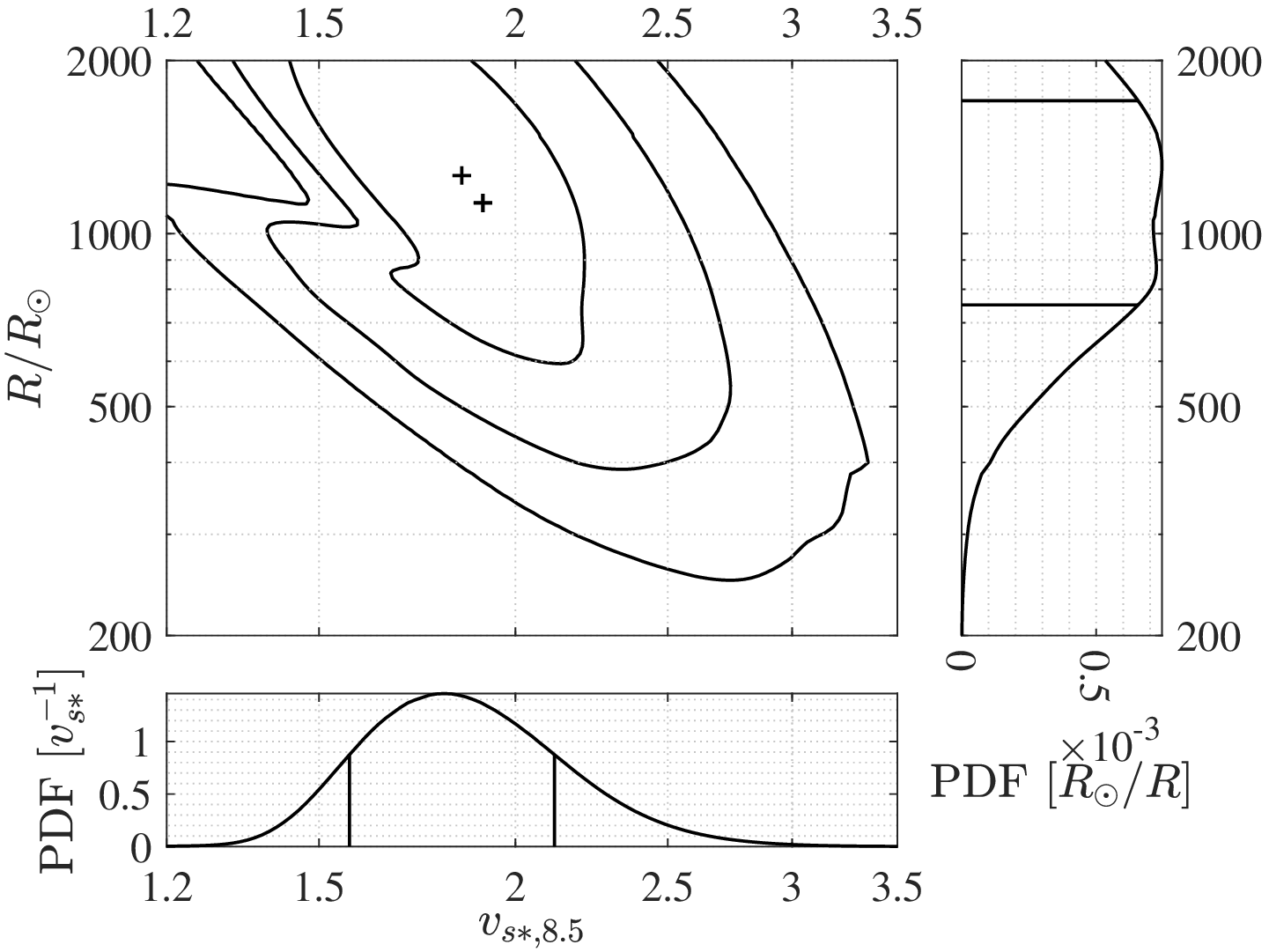}
    \caption{PTF12fkp $R_*$-$v_{s*}$ likelihood map. The solid contour lines represents 1$\sigma$, 2$\sigma$ and 3$\sigma$ of the cumulative likelihood. The plus markers mark the maximal likelihood models. The bottom and right panels show the marginal distributions for $v_{s*}$ and $R_*$.}
    \label{fig:12fkp_CDF}
\end{figure}
\begin{figure}
    \centering
    \includegraphics[width=\linewidth]{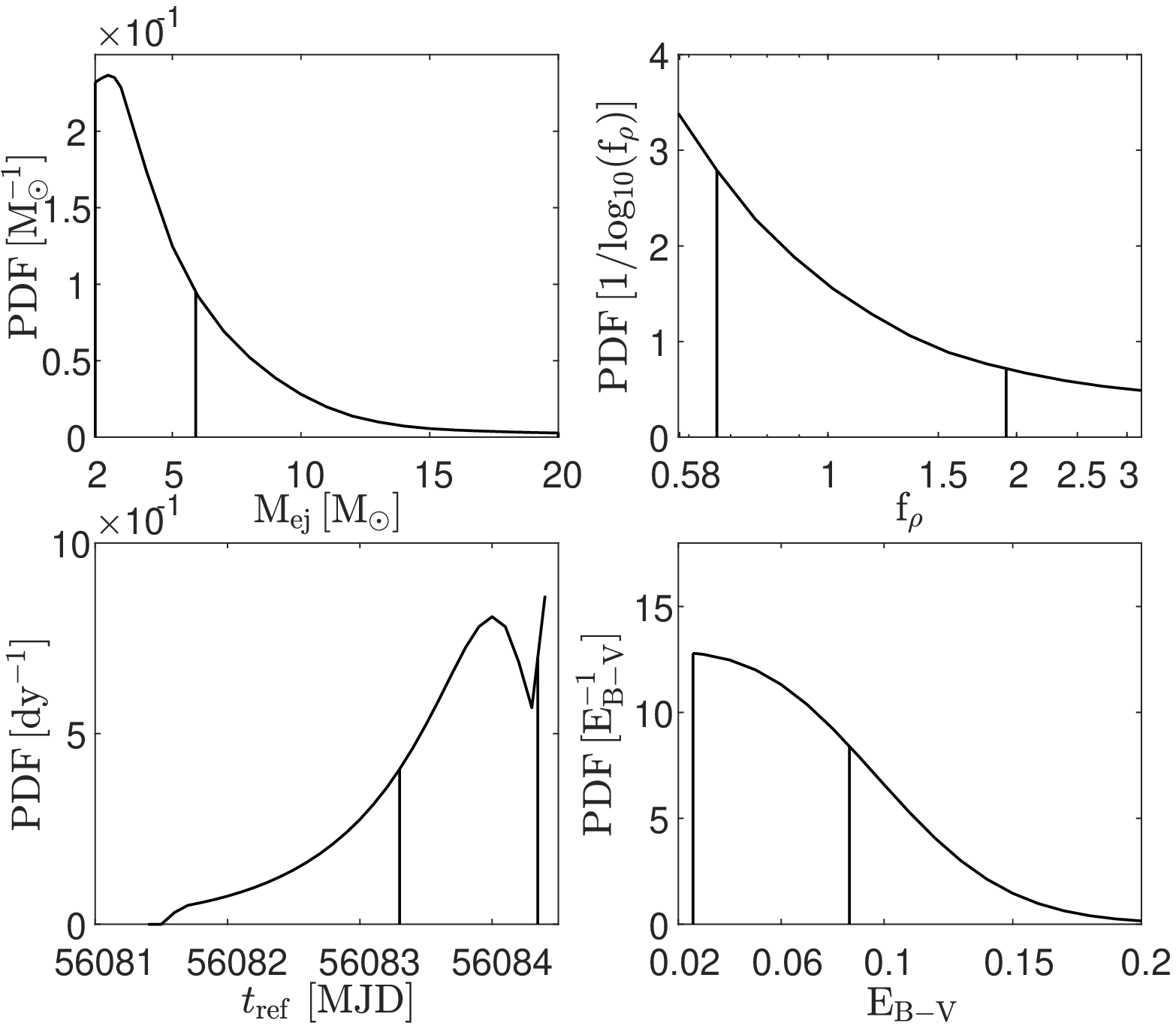}
    \caption{PTF12fkp marginal distribution for $M_{\rm ej}$, $f_\rho$, $t_{\rm ref}$ and $E_{\rm B-V}$.}
    \label{fig:12fkp_Marginals}
\end{figure}

\begin{table}[h]
\caption{PTF12fkp Likelihood local maxima}
\centering
\begin{threeparttable}
\begin{tabular}{l l l }
\hline
\hline
Parameter          & Peak \#1                    & Peak \#2                    \\
\hline
$R_{*}$ [$R_\sun$] & $1262_{-511}^{+441}$        & $1131_{-379}^{+572}$        \\
$v_{s*,8.5}$       & $1.85_{-0.28}^{+0.27}$      & $1.91_{-0.34}^{+0.21}$      \\
$M_{\rm ej}$ [$M_\sun$]&$ 2.1_{- 0.1}^{+ 3.8}$   & $ 2.0_{- 0.0}^{+ 3.9}$      \\
$t_{\rm ref}$ [MJD]& $56084.20_{-0.90}^{+0.14}$  & $56084.11_{-0.81}^{+0.24}$  \\
$f_{\rho}$         & $1.824_{-1.158}^{+0.101}$   & $2.175_{-1.428}^{+0.987}$   \\
$E_{\rm B-V}$      & $0.026_{-0.000}^{+0.061}$   & $0.032_{-0.006}^{+0.055}$   \\
$\chi^2$/dof       &  12.90/ 14                  &  13.06/ 14                  \\
\hline
\end{tabular}
\end{threeparttable}
\label{tab:fkp_results}
\end{table}

PTF12fkp data points together with its most likely models appear in Figure \ref{fig:12fkp_LC}, while the most likely best-fit residuals are presented in Figure \ref{fig:12fkp_Residuals}. The marginalized $R_*$-$v_{s*}$ likelihood map and the marginalized likelihood for each parameter are plotted in Figure \ref{fig:12fkp_CDF} and the marginalized likelihood distributions for $M_{\rm ej}$, $f_\rho$, $t_{\rm ref}$ and $E_{\rm B-V}$ are shown in Figure \ref{fig:12fkp_Marginals}. The local maxima of the likelihood are listed in Table \ref{tab:fkp_results} and marked in Figure \ref{fig:12fkp_Residuals} by plus symbols.

The PTF12fkp $NUV$ data includes a measurement during the rise of the flux to its maximum, and our solutions fit this data point very well (Figure \ref{fig:12fkp_LC}). Its $NUV$ data suffer from a significant loss during the decay of the signal from its peak. Like the other SNe, the marginal progenitor radius likelihood distribution (Fig. \ref{fig:12fkp_CDF}) has non-negligible values for large values of $2000R_\sun$. Unlike the previous SNe, we did not find any dual peak structure in the extinction marginalized likelihood distributions (Fig. \ref{fig:12fkp_Marginals}) which allow us to limit our prior and to suppress the likelihood of these large radius models. While the solutions we find (Tab. \ref{tab:fkp_results}) fit the data very well, the limits we receive for the radius are very wide. If some of the $NUV$ data points were not lost, we may have had a better constraints on the radius.

\subsection{PTF12ftc}

\begin{figure}
    \centering
    \includegraphics[width=\linewidth]{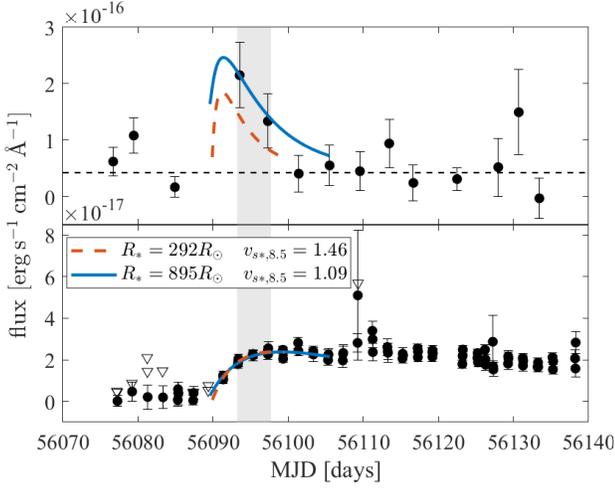}
    \caption{PTF12ftc data points and best fitting models. The upper panel describes the $NUV$ band data and the lower one is for the $R$ band. The dashed horizontal line in the top panel represents measured $NUV$ background. The triangles in the bottom stand for $3\sigma$ limits. The colored lines show the different solutions. The $R_*$ and $v_{s*}$ values of these solution are listed in the legend, while the other parameters are in Table \ref{tab:ftc_results}. The grayed background area marks the $NUV$ transient. Data points external to this area were used to calculate the $NUV$ background.}
    \label{fig:12ftc_LC}
\end{figure}

\begin{figure}
    \centering
    \includegraphics[width=\linewidth]{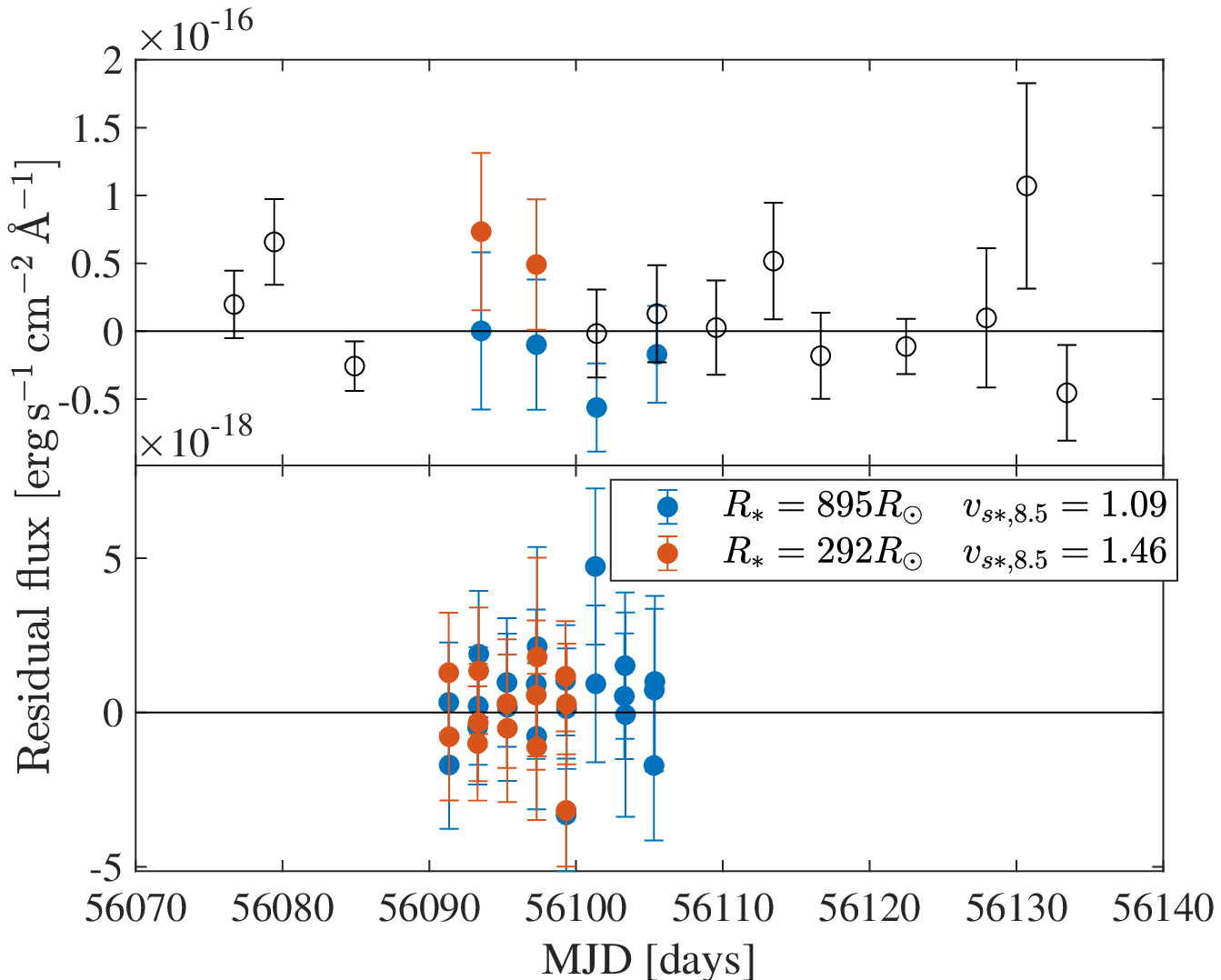}
    \caption{PTF12ftc best-fits residuals for the models plotted on Fig. \ref{fig:12ftc_LC}. The top panel shows the $NUV$ band residuals and the bottom panel the $R$ band ones. The color code is the same as in Fig. \ref{fig:12ftc_LC}. The empty black circles are the residuals of the $NUV$ background estimation.}
    \label{fig:12ftc_Residuals}
\end{figure}

\begin{figure}
    \centering
    \includegraphics[width=\linewidth]{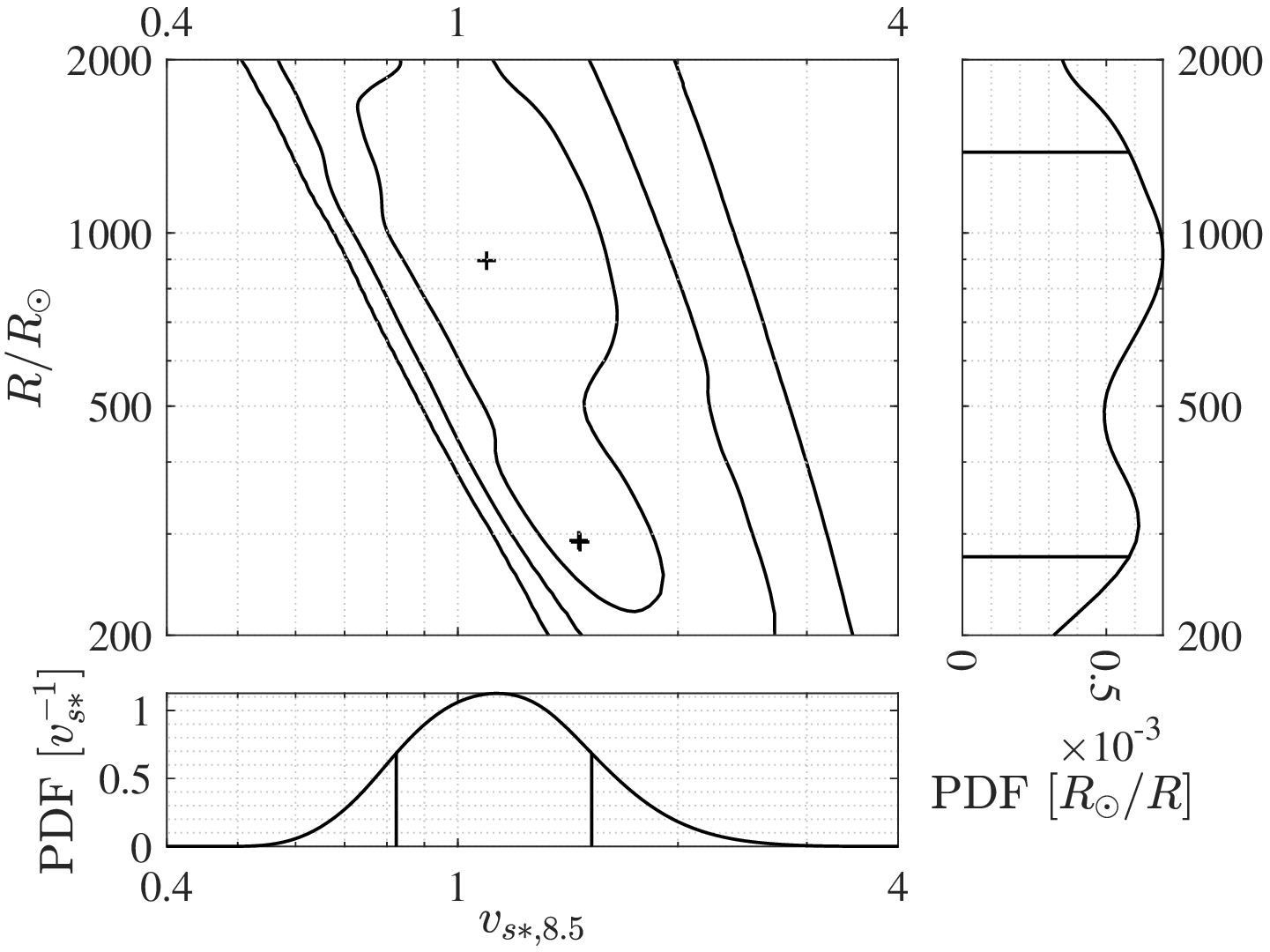}
    \caption{PTF12ftc $R_*$-$v_{s*}$ likelihood map. The solid contour lines represents 1$\sigma$, 2$\sigma$ and 3$\sigma$ of the cumulative likelihood. The plus markers indicate the maximal likelihood solutions. The bottom and right panels show the marginal distributions for $v_{s*}$ and $R_*$.}
    \label{fig:12ftc_CDF}
\end{figure}
\begin{figure}
    \centering
    \includegraphics[width=\linewidth]{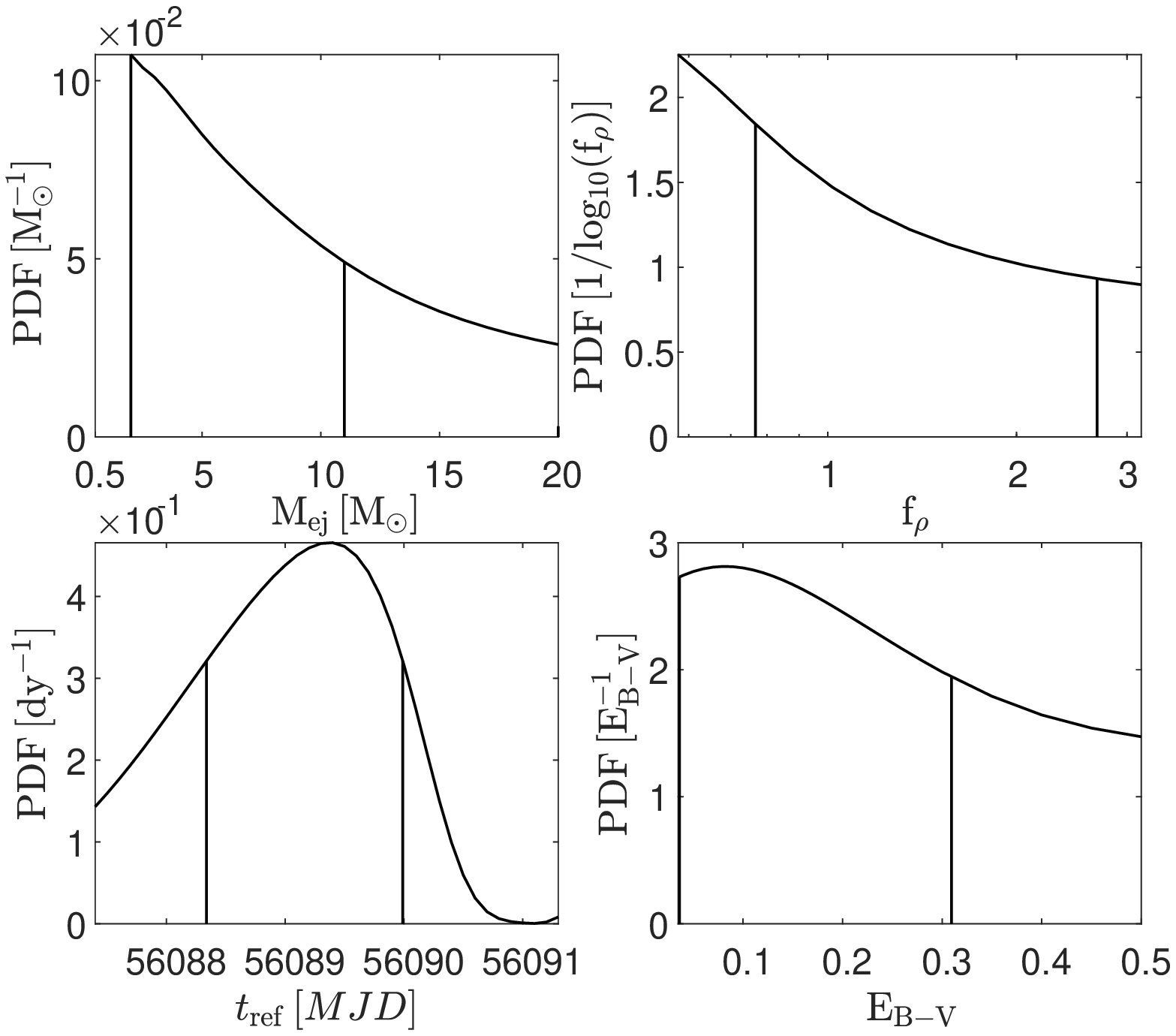}
    \caption{PTF12ftc marginal distributions for $M_{\rm ej}$, $f_\rho$, $t_{\rm ref}$ and $E_{\rm B-V}$.}
    \label{fig:12ftc_Marginals}
\end{figure}

\begin{table}[t]
\caption{PTF12ftc Likelihood local maxima}
\centering
\begin{threeparttable}
\begin{tabular*}{0.99\linewidth}{l l l l}
\hline
\hline
Parameter          & Peak \#1\tnote{a}           & Peak \#2\tnote{a}           & Peak \#3                    \\
\hline
$R_{*}$ [$R_\sun$] & $ 292_{- 19}^{+1090}$       & $ 290_{- 17}^{+1092}$       & $ 895_{-621}^{+487}$        \\
$v_{s*,8.5}$       & $1.46_{-0.64}^{+0.06}$      & $1.47_{-0.65}^{+0.05}$      & $1.09_{-0.27}^{+0.43}$      \\
$M_{\rm ej}$ [$M_\sun$]& $ 4.4_{- 2.4}^{+ 6.6}$  & $ 4.9_{- 2.9}^{+ 6.1}$      & $ 2.0_{- 0.0}^{+ 9.0}$      \\
$t_{\rm ref}$ [MJD]& $56089.86_{-1.52}^{+0.13}$  & $56089.85_{-1.51}^{+0.14}$  & $56089.49_{-1.15}^{+0.50}$  \\
$f_{\rho}$         & $2.229_{-1.506}^{+0.153}$   & $1.867_{-1.202}^{+0.157}$   & $2.395_{-1.628}^{+0.293}$   \\
$E_{\rm B-V}$      & $0.036_{-0.000}^{+0.273}$   & $0.036_{-0.000}^{+0.273}$   & $0.038_{-0.001}^{+0.272}$   \\
$\chi^2$/dof       &   4.62/  6                  &   4.63/  6                  &  13.00/ 15                  \\
\hline
\end{tabular*}
\begin{tablenotes}
            \item[a] These solutions have a poor match to the NUV data points (see Figs. \ref{fig:12ftc_LC} and \ref{fig:12ftc_Residuals}) and should be rejected.
\end{tablenotes}
\end{threeparttable}
\label{tab:ftc_results}
\end{table}

PTF12ftc data points together with its most likely models appear in Figure \ref{fig:12ftc_LC}, while the most likely best-fit residuals are presented in Figure \ref{fig:12ftc_Residuals}. The marginalized $R_*$-$v_{s*}$ likelihood map and the marginalized likelihood for each parameter are plotted in Figure \ref{fig:12ftc_CDF} and the marginalized likelihood distributions $M_{\rm ej}$, $f_\rho$, $t_{\rm ref}$ and $E_{\rm B-V}$ are shown in Figure \ref{fig:12ftc_Marginals}. The local maxima of the likelihood are listed in Table \ref{tab:ftc_results} and marked on Figure \ref{fig:12ftc_Residuals} by plus symbols.

While all the three solution in Table \ref{tab:ftc_results} are in good agreement with the $R$-band readings, the solutions with the lower progenitor radius ($\sim300R_\sun$) do not match the $NUV$ data points (see Figs. \ref{fig:12ftc_LC} and \ref{fig:12ftc_Residuals} for a comparison of the solutions against the data). This anomaly is a result of the small amount of data points these solution are valid for. In addition these solutions are dominated by the $R$-band points (these solutions are valid for two $NUV$-band points and ten $R$-band points). The good match to the $R$-band points covers for the mismatch with the $NUV$ points, leading to an overall reasonable goodness of fit score. However, since we require a solution to match well all the different bands in addition to its overall goodness of fit score, we reject these solutions. We report them since our \verb|SOPRANOS| tool found them as possible solutions.
The $NUV$ data of this SN suffers from a data loss between the last non-detection and the first detection of the SN. The marginal distribution for the extinction has a non-negligible likelihood even for very high values of $E_{B-V}$. The progenitor radius for this SN has poor limits which may be a result of the small amount of $NUV$ data points the solutions are valid for. Examining the marginal distributions for the other model parameters (Fig. \ref{fig:12ftc_Marginals}) we did not found any secondary peaks which may have a correlation with the $\gtrsim1000R_\sun$ progenitor radius models. If we have found such a correlation we could narrow down our prior to avoid the non-physical large progenitor radius values. If we set a requirement of a minimal number of 13 valid data points (7 degrees of freedom), the low radii solutions  ($\sim300R_\sun$) disappear, leading to a tighter progenitor radius limit.

\subsection{PTF12fhz}
\begin{figure}
    \centering
    \includegraphics[width=\linewidth]{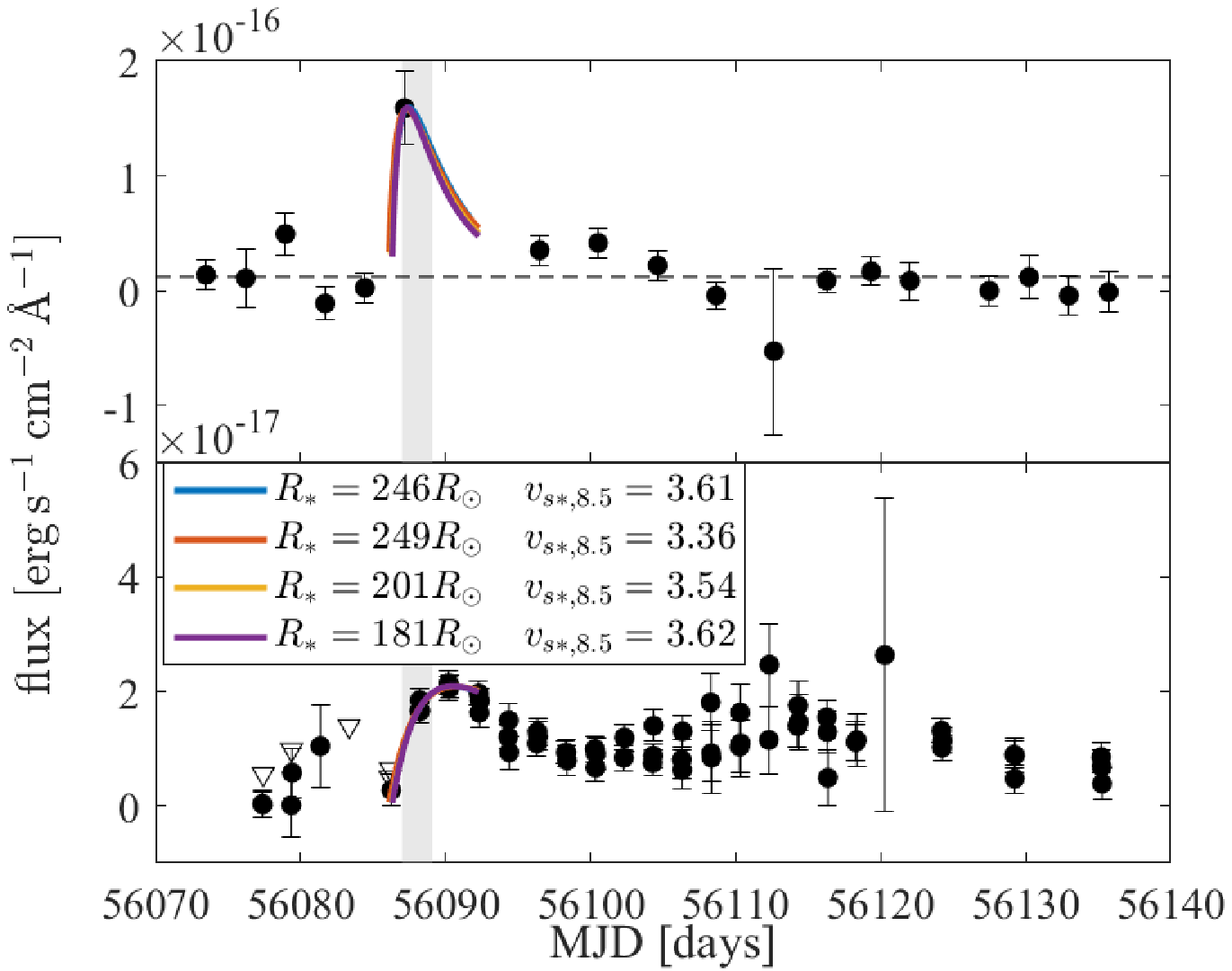}
    \caption{PTF12fhz data points and best-fitting models. The upper panel describes the $NUV$ band data and the lower one is for the $R$ band. The dashed horizontal line in the top panel represents the measured $NUV$ background. The triangles in the bottom panel stand for $3\sigma$ limits. The colored lines show the different solutions. The $R_*$ and $v_{s*}$ values for these solutions are listed in the legend, while the other parameters are in Table \ref{tab:fhz_results}. The grayed background area marks the $NUV$ transient. Data points external to this area were used to calculate the $NUV$ background.}
    \label{fig:12fhz_LC}
\end{figure}

\begin{figure}
    \centering
    \includegraphics[width=\linewidth]{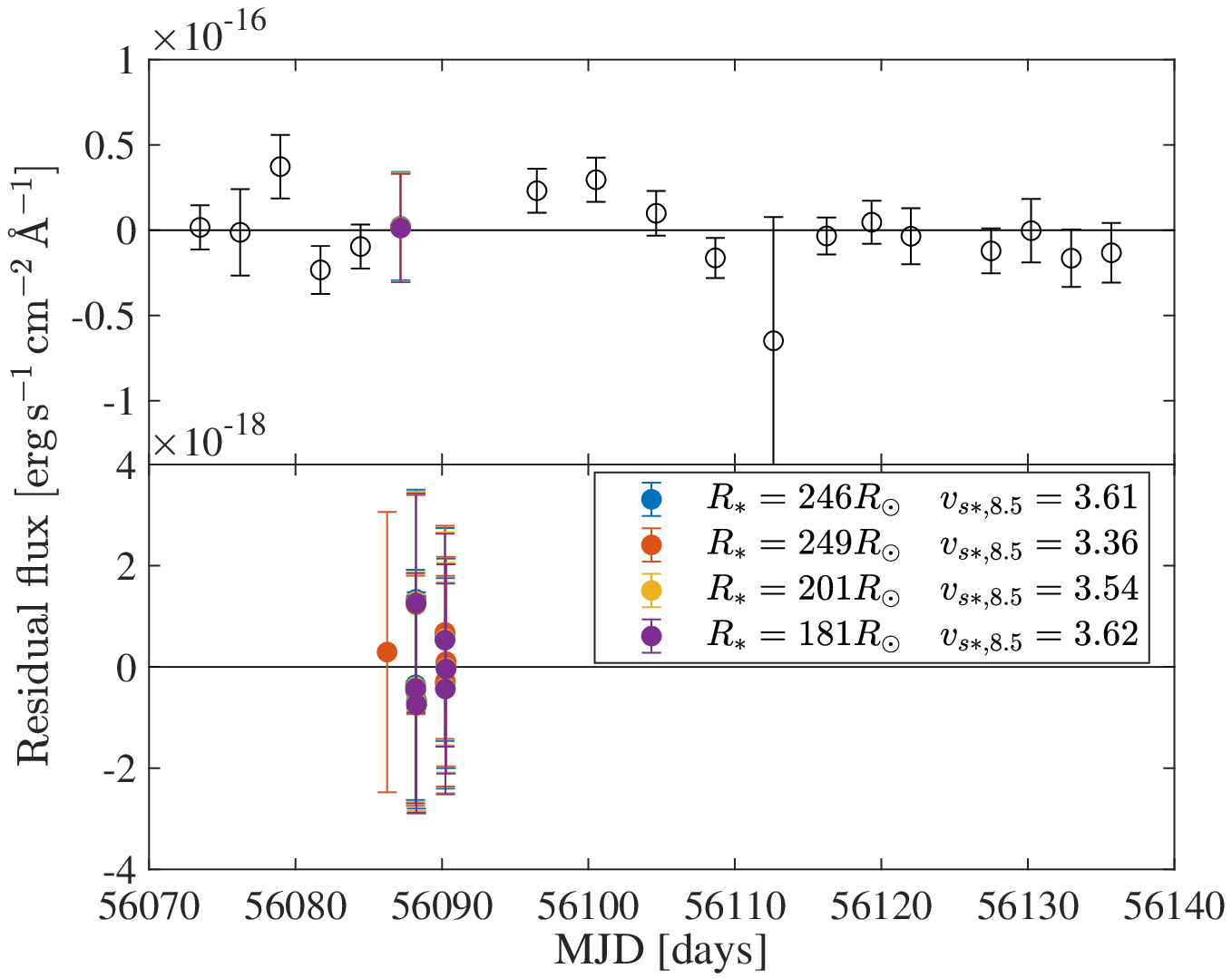}
    \caption{PTF12fhz best-fits residuals for the models plotted on Fig. \ref{fig:12fhz_LC}. The upper panel shows the $NUV$ band residuals and the bottom panel the $R$ band. The color code is the same as in Fig. \ref{fig:12fhz_LC}. The empty black circles are the residuals of the $NUV$ background estimation.}
    \label{fig:12fhz_Residuals}
\end{figure}

\begin{figure}
    \centering
    \includegraphics[width=\linewidth]{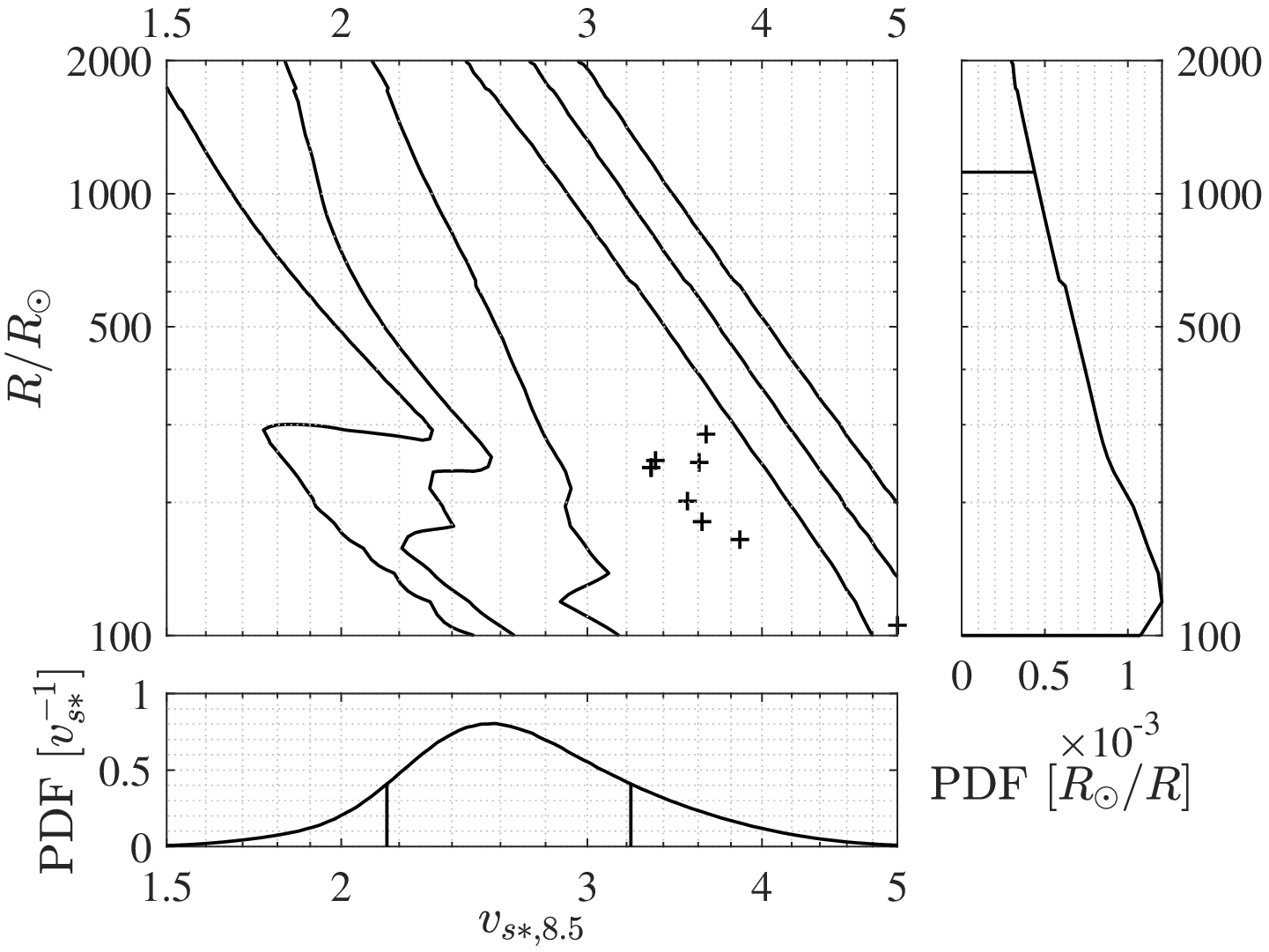}
    \caption{PTF12fhz $R_*$-$v_{s*}$ likelihood map. The solid contour lines represents 1$\sigma$, 2$\sigma$ and 3$\sigma$ of the cumulative likelihood. The plus markers indicate the maximal likelihood models. The bottom and right panels show the marginal distributions for $v_s*$ and $R_*$.}
    \label{fig:12fhz_CDF}
\end{figure}
\begin{figure}
    \centering
    \includegraphics[width=\linewidth]{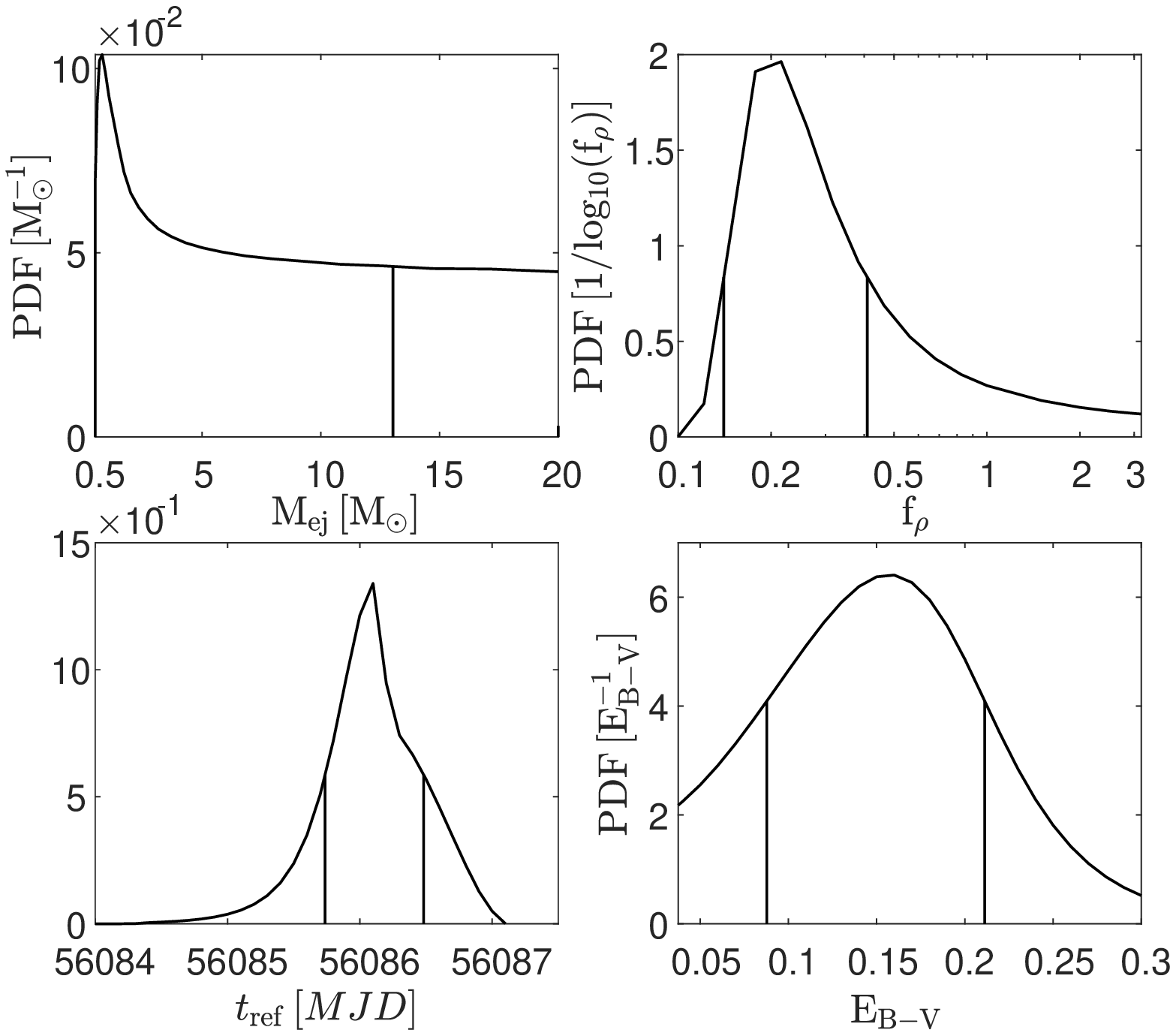}
    \caption{PTF12fhz marginal distributions for $M_{\rm ej}$, $f_\rho$, $t_{\rm ref}$ and $E_{\rm B-V}$.}
    \label{fig:12fhz_Marginals}
\end{figure}

\begin{table}[t]
\caption{PTF12fhz Likelihood local maxima}
\centering
\begin{threeparttable}
\begin{tabular*}{\linewidth}{l l l l}
\hline
\hline
Parameter          & Peak \#1                    & Peak \#2                    & Peak \#3                    \\
\hline
$R_{*}$ [$R_\sun$] & $ 246_{-146}^{+872}$        & $ 201_{-101}^{+917}$        & $ 181_{- 81}^{+937}$        \\
$v_{s*,8.5}$       & $3.61_{-1.24}^{+0.00}$      & $3.54_{-1.19}^{+0.00}$      & $3.62_{-1.25}^{+0.00}$      \\
$M_{\rm ej}$ [$M_\sun$]& $ 1.5_{- 1.0}^{+11.5}$  & $ 4.7_{- 4.2}^{+ 8.4}$      & $ 5.7_{- 5.2}^{+ 7.4}$      \\
$t_{\rm ref}$ [MJD]& $56086.23_{-0.50}^{+0.25}$  & $56086.24_{-0.50}^{+0.25}$  & $56086.29_{-0.56}^{+0.19}$  \\
$f_{\rho}$         & $0.902_{-0.716}^{+0.000}$   & $0.447_{-0.297}^{+0.000}$   & $0.401_{-0.260}^{+0.009}$   \\
$E_{\rm B-V}$      & $0.098_{-0.010}^{+0.114}$   & $0.082_{-0.000}^{+0.124}$   & $0.077_{-0.000}^{+0.125}$   \\
$\chi^2$/dof       &   0.60/  1                  &   0.60/  1                  &   0.60/  1                  \\
\hline
\hline
                   & Peak \#4                    & Peak \#5                    & Peak \#6                    \\
\hline
$R_{*}$ [$R_\sun$] & $ 165_{- 65}^{+954}$        & $240_{-140}^{+878}$         & $ 249_{-149}^{+869}$        \\
$v_{s*,8.5}$       & $3.86_{-1.42}^{+0.00}$      & $3.33_{-1.09}^{+0.00}$      & $3.36_{-1.10}^{+0.00}$      \\
$M_{\rm ej}$ [$M_\sun$]&$ 5.1_{- 4.6}^{+ 7.9}$   & $10.3_{- 9.8}^{+ 2.8}$      & $ 2.5_{- 2.0}^{+10.6}$      \\
$t_{\rm ref}$ [MJD]& $56086.40_{-0.67}^{+0.08}$  & $56086.23_{-0.50}^{+0.25}$  & $56086.06_{-0.33}^{+0.42}$  \\
$f_{\rho}$         & $0.403_{-0.262}^{+0.007}$   & $0.271_{-0.131}^{+0.139}$   & $0.684_{-0.507}^{+0.000}$   \\
$E_{\rm B-V}$      & $0.077_{-0.000}^{+0.125}$   & $0.087_{-0.000}^{+0.123}$   & $0.089_{-0.002}^{+0.122}$   \\
$\chi^2$/dof       &   0.60/  1                  &   0.61/  1                  &   0.61/  2                  \\
\hline
\hline
                   & Peak \#7                    & Peak \#8                   \\
\hline
$R_{*}$ [$R_\sun$] & $ 286_{-186}^{+833}$        & $ 105_{-  5}^{+1013}$      \\
$v_{s*,8.5}$       & $3.65_{-1.27}^{+0.00}$      & $5.00_{-2.48}^{+0.00}$     \\
$M_{\rm ej}$ [$M_\sun$]&$ 7.3_{- 6.8}^{+ 5.7}$   & $ 6.4_{- 5.9}^{+ 6.6}$     \\
$t_{\rm ref}$ [MJD]& $56086.12_{-0.39}^{+0.36}$  & $56086.13_{-0.40}^{+0.35}$ \\
$f_{\rho}$         & $0.362_{-0.222}^{+0.047}$   & $0.483_{-0.328}^{+0.000}$  \\
$E_{\rm B-V}$      & $0.198_{-0.110}^{+0.013}$   & $0.234_{-0.129}^{+0.000}$  \\
$\chi^2$/dof       &   4.68/  3                  &  10.54/  5                 \\
\hline
\end{tabular*}
\end{threeparttable}
\label{tab:fhz_results}
\end{table}
PTF12fhz data points together with its most likely models appear in Figure \ref{fig:12fhz_LC}, while the most likely best-fit residuals are presented in Figure \ref{fig:12fhz_Residuals}. The marginalized $R_*$-$v_{s*}$ likelihood map and the marginalized likelihood for each parameter are plotted in Figure \ref{fig:12fhz_CDF} and the marginalized likelihood distributions $M_{\rm ej}$, $f_\rho$, $t_{\rm ref}$ and $E_{\rm B-V}$ are shown in Figure \ref{fig:12fhz_Marginals}. The local maxima of the likelihood are listed in Table \ref{tab:fhz_results} and marked on Figure \ref{fig:12fhz_Residuals} by plus symbols.

The $R$-band light curve (Figure \ref{fig:12fhz_LC}) shows a dual peak structure, which is common to type IIb SNe, like PTF12fhz. SW17 have shown that their model can explain the first peak of the two when extending the $f_\rho$ prior to values smaller than $\sqrt{1/3}$. Following SW17, as mentioned in \S\ref{subsec:prior}, we extended our $f_\rho$ prior to include values starting from 0.1. Like the \citet{SapirWaxman2017} results for LSQ14bdq and 1993J, our solutions for PTF12fhz converge to low radius progenitor $R_*<300R_\sun$ and low $f_\rho$ value. Here we show that the model also matches the $NUV$ band data. Such low radius solutions are valid only for a short period of time, and our low cadence survey has obtained only a small number of data points during the models validity period. Our most likely models are valid for small number of 7-11 data points in the two bands together. While the solutions have a good goodness of fit, the very small number of data points they are valid for makes the statistical significance of this result poor. We cannot place a tight limit on the progenitor radius. Its marginal likelihood distribution decays slowly towards $2000R_\sun$ and beyond. This result may be explained by the fact that we have only one valid $NUV$ point, and few data points in epochs following it were lost. The ejecta mass marginal likelihood distribution has a peak at $2M_\sun$ and then has a non-zero asymptotic value (Figure \ref{fig:12fhz_Marginals}). This can be explained by our solutions having small $f_\rho$ values and that Eq. \ref{eq:RW11} depends on the multiplication of $f_\rho$ and $M_{\rm ej}$. For a small $f_\rho$ value, a small absolute change is a large relative change. In order to maintain the value of the multiplication of the two, $M_{\rm ej}$ value should be changed by an inverse factor, which leads to large change of its absolute value. While the different solutions range $M_{\rm ej}=1.5-10.3M_\sun$ and $f_\rho=0.27-0.9$ their multiplication is in the limited range of $f_\rho M_{\rm ej}=1.4-2.7M_\sun$.

\section{Discussion}\label{s:Discussion}
A phenomenon we encountered which is of general relevance is that when we fit SW17 models to the data, we find a preference for large progenitor radii ($\gtrsim1000R_\sun$). The reason for this preference is that large radius models become valid at late times (Equation \ref{eq:t_min}), typically after the NUV signal is no longer significant. When we use the MSW20 extension to the SW17 model, the early $NUV$ data points became valid for all models, and we find that progenitor radii of few hundreds of solar radii have higher likelihood than the $R_*\gtrsim1000R_\sun$ solutions. This is important to account for when applying either the RW11 or the SW17 models to observations.

The marginalized progenitor-radius likelihood distributions of all the analyzed SNe have a non-negligible likelihood for presumably non-physical progenitor radii of about $2000R_{\sun}$. For PTF12ffs and PTF12gnt, where we have many data points in both $NUV$ and $R$-bands we have recognized a correlation between theses large radius solutions and a secondary peak in the  marginalized extinction distribution, with a higher extinction value. This is explained by Equations \ref{eq:RW11} and \ref{eq:L_SKW}. When we limit the prior on the extinction values to include only the peak with lower extinction values for those SNe, the high radius solutions disappear. For the other SNe, for which we obtained fewer data points and only a few $NUV$ data points with values larger than their $NUV$ background, we do not identify a secondary peak in any of the other parameters marginalized distributions. This behaviour may be explained by the small number of data points within the model validity period and the characteristics of the $\chi^2$ probability density function for small number of degrees of freedom, which does not fall sharply from maximum at $\chi^2\sim\nu$, resulting in a slow decay of the likelihood function from its maximum. Our work demonstrates the importance of constraining the extinction towards these events in order to properly derive their progenitor radii. The largest difference between the predicted light curves of best-fit models for PTF12gnt and PTF12ffs is at times where we do not have an $NUV$-band measurement. We believe that if we had a higher cadence in this band, data points in these times could differentiate better between the best-fit models, leading to a tighter constraint.

Whether or not we were able to achieve a tight constraint on the progenitor radius, all the solutions \verb|SOPRANOS| converged to have an acceptable goodness of fit with the exception of PTF12gnt. For PTF12gnt we identify a group of three $R$-band measurements on the first night of the SN detection which do not match any model. While we do not have a statistical justification to ignore those data points, when we ignore them the solutions match the data well, and also suppress the likelihood of the nonphysical large radius solutions.

The light curves of PTF12gnt, PTF12fkp, and PTF12ftc demonstrate that the $NUV$-band data points are critical for narrowing the confidence intervals of the solutions. The double peaked SN PTF12fhz demonstrates that SW17 models are able to explain not only the first peak of the double peaked $R$-band light curve, but also its $NUV$ behaviour.

We conclude the discussion by the fact that higher cadence $NUV$ surveys, with higher measurement accuracy as expected to be conducted by the \textit{ULTRASAT} space mission, would allow us to obtain a tight constraints on SN progenitor radii. This fact was shown by analysis in \citet{RubinEtAl2017}.

\section{Conclusion}\label{S:Conculusion}
We have developed \verb|SOPRANOS|, a maximum likelihood fitting tool, which takes into account all the valid data points for each shock-cooling model and uses the likelihood function to compare between the different models. This is in contrast to previous works which used a constant set of data points for all the models, ignoring the validity time range of the models.

We have analyzed the SNe detected during the GALEX/PTF experiment using \verb|SOPRANOS|, and found a good agreement between the MSW model and the data. The introduction of the MSW model extension allowed us to utilize all the early $NUV$ data points. For two SNe with dense $NUV$ data points we also achieved a good constraint on the progenitor radius. This constraint was achieved despite the low cadence of our survey. Higher cadence $NUV$ surveys, such as \textit{ULTRASAT}, will provide definitive measurements of the progenitor radii of core-collapse SNe.

We also have demonstrated that supergiant stars with small envelope to core ratio $f_\rho$ may explain the double peaked type IIb SNe, as shown by \citet{SapirWaxman2017}, and that their solution is also compatible with the $NUV$ band data points of PTF12fhz, complementing the visible light data analyzed by these authors.

\acknowledgments
We want to thank Barak Zackay for useful discussions leading to the development of \texttt{SOPRANOS} formalism.

E.O.O. is grateful for the support by grants from the Israel Science Foundation, Minerva, Israeli Ministry of Technology and Science, the US-Israel Binational Science Foundation, Weizmann-UK, Weizmann-Yale, the Weizmann-Caltech grants, the Norman E. Alexander Family M. Foundation ULTRASAT Data Center Fund, Jonathan Beare, Andr\'e Deloro Institute for Space and Optics Research, Schwartz/Reisman Collaborative Science Program and the Willner Family Leadership Institute for the Weizmann Institute of Science.

AGY’s research is supported by the EU via ERC grant No. 725161, the ISF GW excellence center, an IMOS space infrastructure grant and BSF/Transformative and GIF grants, as well as The Benoziyo Endowment Fund for the Advancement of Science, the Deloro Institute for Advanced Research in Space and Optics, The Veronika A. Rabl Physics Discretionary Fund, Minerva, Yeda-Sela and the Schwartz/Reisman Collaborative Science Program; AGY is the recipient of the Helen and Martin Kimmel Award for Innovative Investigation.

MMK acknowledges generous support from the David and Lucille Packard Foundation.

\bibliography{ref}

\end{document}